\documentclass[twocolumn]{aastex631}

\usepackage{booktabs}
\usepackage{gensymb}
\usepackage{upgreek}
\usepackage{multirow}
\usepackage{bigdelim}
\usepackage{hyperref}
\usepackage[graphicx]{realboxes}

\newcommand{\kmss}{km~s$^{-1}\ $}
\newcommand{\Ms}{$M_{\odot}\ $}
\newcommand{\Lyas}{Ly$\alpha$ }
\newcommand{\HIs}{\ion{H}{1} }
\newcommand{\OIs}{\ion{O}{1} }
\newcommand{\SiIIs}{\ion{Si}{2} }
\newcommand{\SiIIIs}{\ion{Si}{3} }
\newcommand{\SiIVs}{\ion{Si}{4} }
\newcommand{\CIIs}{\ion{C}{2} }
\newcommand{\CIVs}{\ion{C}{4} }
\newcommand{\NVs}{\ion{N}{5} }
\newcommand{\MgIIs}{\ion{Mg}{2} }
\newcommand{\sigls}{$\sigma_\mathrm{LOS}\ $}

\newcommand{\M}{$M_{\odot}$}
\newcommand{\Lya}{Ly$\alpha$}
\newcommand{\HI}{\ion{H}{1}}
\newcommand{\OI}{\ion{O}{1}}
\newcommand{\SiII}{\ion{Si}{2}}
\newcommand{\SiIII}{\ion{Si}{3}}
\newcommand{\SiIV}{\ion{Si}{4}}
\newcommand{\CII}{\ion{C}{2}}
\newcommand{\CIV}{\ion{C}{4}}
\newcommand{\NV}{\ion{N}{5}}

\newcommand{\kms}{km~s$^{-1}$}

\begin{document}

\title{DIISC-VI (COS-DIISC): UV Metal Absorption Relative to the H I disk of Galaxies}

\shorttitle{DIISC-VI: CGM Metal Features}
\shortauthors{Koplitz et al.}

\correspondingauthor{Brad Koplitz} 
\email{Brad.Koplitz@asu.edu}

\author[0000-0001-5530-2872]{Brad Koplitz}
\affil{School of Earth \& Space Exploration, 
Arizona State University, 781 Terrace Mall, Tempe, AZ 85287, USA}

\author[0000-0002-2724-8298]{Sanchayeeta Borthakur}
\affil{School of Earth \& Space Exploration, 
Arizona State University, 781 Terrace Mall, Tempe, AZ 85287, USA}

\author[0000-0001-6670-6370]{Timothy Heckman}
\affil{School of Earth \& Space Exploration, 
Arizona State University, 781 Terrace Mall, Tempe, AZ 85287, USA}
\affil{Department of Physics \& Astronomy, 
Johns Hopkins University, Bloomberg Centre, 
3400 N. Charles Street, Baltimore, MD 21218, USA}

\author[0000-0002-3472-0490]{Mansi Padave}
\affil{School of Earth \& Space Exploration, 
Arizona State University, 781 Terrace Mall, Tempe, AZ 85287, USA}

\author[0000-0002-5506-3880]{Tyler McCabe}
\affil{School of Earth \& Space Exploration, 
Arizona State University, 781 Terrace Mall, Tempe, AZ 85287, USA}

\author[0000-0002-7982-412X]{Jason Tumlinson}
\affil{Space Telescope Science Institute, 
3700 San Martin Drive, Baltimore, MD, USA}

\author[0000-0003-0724-4115]{Andrew J. Fox}
\affil{AURA for ESA, Space Telescope Science Institute, 
3700 San Martin Drive, Baltimore, MD 21218, USA}
\affil{Department of Physics \& Astronomy, Johns Hopkins University, 
3400 N. Charles Street, Baltimore, MD 21218, USA}

\author{Guinevere Kauffmann}
\affil{Max-Planck-Institute f{\"u}r Astrophysik, Karl-Schwarzschild-Str 1, D-85740 Garching, Germany}

\begin{abstract}

As part of the Deciphering the Interplay between the Interstellar medium, Stars, and the Circumgalactic medium (DIISC) survey, we present the UV metal absorption features in the circumgalactic medium (CGM) near the \ion{H}{1} gas disk ($<$4.5$R_\mathrm{HI}$) of 31 nearby galaxies through quasar absorption line spectroscopy.
Of the ions under study, \ion{Si}{3} $\lambda1206$ was most frequently detected (18 of 31 sight lines), while \ion{C}{2} $\lambda1334$ and \ion{Si}{2} $\lambda1260$ were detected in 17 and 15 of 31 sight lines, respectively.
Many components were consistent with photoionization equilibrium models; most of the cold and cool gas phase clouds were found to have lengths smaller than 2~kpc.
Sight lines with smaller impact parameters ($\rho$) normalized by the galaxy's virial radius ($R_\mathrm{vir}$) and \ion{H}{1} radius ($R_\mathrm{HI}$) tend to have more components and larger rest-frame equivalent widths ($W_r$) than those that probe the CGM at larger radii.
In particular, we find that the location of metals are better traced by $\rho$ / $R_\mathrm{HI}$ rather than the traditional $\rho$ / $R_\mathrm{vir}$.
Larger covering fractions are found closer to galaxies, with a radial decline that depends on the $W_r$ limit used.
Our results provide new insights into the spatial distribution of metals around the \ion{H}{1} disks of low-redshift galaxies.

\end{abstract}


\section{INTRODUCTION}

The halo gas associated with galaxies extends well beyond where we see stars.
It is known that this medium, referred to as the circumgalactic medium (CGM), is multiphase and can be found out to and beyond the virial radius of galaxies ($R_\mathrm{vir}$;~\citealt{Tumlinson2017} and references therein).
This gas is fundamental to understanding how galaxies evolve as it is a major component of their baryon cycles \citep{PerouxHowk2020,DonahueVoit2022}.
Star formation is thought to be triggered from gas accreting from the CGM and the intergalactic medium (e.g.,~\citealt{Keres2005,Dekel2006,Dekel2009,Somerville2015,Hafen2022,Decataldo2023}) while feedback mechanisms, such as stellar winds, return this gas back to the CGM (e.g.,~\citealt{Veilleux2005,Shen2012,Suresh2015,Faucher2016,Kauffmann2016,Machado2018,Davies2020,Sorini2020,Voit2020,Zinger2020}).

There are numerous benefits to studying the metal absorbers in the CGM.
They often show multiple components even when the associated \Lyas is saturated or damped (e.g.,~\citealt{Chen2002,Werk2013,Bordoloi2014,Borthakur2016,Keeney2017,Muzahid2018}), allowing the column densities and Doppler widths of absorbers to be measured more precisely.
Metals also probe the multiple phases of the CGM (e.g.,~\citealt{Werk2013,Haislmaier2021,Sameer2024}), enabling the ionization state around galaxies to be constrained.
Additionally, the physical size of these CGM absorbing clouds can be estimated by comparing ionization models to observed metal lines (e.g.,~\citealt{Stocke2013,Werk2014,Muzahid2018}).

The Cosmic Origins Spectrograph (COS)-Halos survey \citep{Tumlinson2013} studied the CGM around 44 $L_\star$ galaxies, both passive and star-forming, at redshifts of $z \approx 0.2$.
In particular, at least one line from intermediate ions was found in 75\% of systems.
This suggested the presence of a cool ($T \sim 10^{4}-10^{5}$~K), metal-rich CGM component around most galaxies \citep{Werk2013}.
The strength of the metal absorbers was found to decrease with increasing impact parameter ($\rho$), implying a decreasing metal surface density profile.
This radial decline has been reproduced by simulations (e.g., \citealt{Appleby2021}).
Both star-forming and quenched galaxies were found to have CGM masses of $\sim$$10^{10}$~$M_\odot$, meaning the early types have less CGM mass relative to their stellar and halo masses as compared to star-forming galaxies, which may be a signature of quenching \citep{Thom2012}.
Additionally, the quenching of star formation was found to take place regardless of the presence of strong \HIs absorption, contrary to some theoretical expectations \citep{Thom2012}.
However, \citet{Burchett2018} found that the CGM of X-ray bright clusters contained less \HIs relative to field galaxies, which may be indicative of quenching.
More recently, \citet{Tch2023} found that star-forming galaxies tend to have more \ion{O}{6} in their CGM than passive galaxies at the same stellar mass.

Rather than statistically studying many CGMs probed by single QSOs, \citet{Keeney2013} investigated a single galaxy (ESO~157-49) using three sight lines, a subsample of 11 sight lines in the catalog of \citet{Stocke2013}.
The two sight lines probing along the disk of the edge-on galaxy contained absorption from \SiIII, \CIV, and \SiIV.
These absorbers provided further evidence that the warm phase of the CGM have high covering fractions, similar to what was found by \citet{Tumlinson2011b}.
The sight line near the galaxy's minor axis only showed detection of \HI; however, it is worth noting that this sight line is located at nearly twice the virial radius of the galaxy.
Using these sight lines, along with the full sample from \citet{Stocke2013} and a subsample of the \citet{DanforthShull2008} catalog, \citet{Keeney2017} found that the CGM of star-forming galaxies likely contains a few thousand cool clouds with sizes between 1 and 20~kpc, with many more at sizes between 200~pc and 1~kpc.
These cloud sizes, including some as small as 10~pc, have been found by \citet{Rigby2002} and the Cosmic Ultraviolet Baryon Survey (CUBS; \citealt{Zahedy2021}).

In a similar study, \citet{Bowen2016} used four QSO sight lines to study the CGM of NGC 1097.
The three sight lines closest to the galaxy contained \SiIIIs while \SiII, \CII, and \SiIVs were only found along the closest sight line.
The dearth of \SiIVs detections suggests that the warm phase may not be as prevelent as was found by \citet{Tumlinson2011b} and \citet{Keeney2013}.

The GALEX Arecibo SDSS Survey (GASS) looked at \Lyas absorption features in the outer CGM of 36 far-ultraviolet (FUV) bright QSO-galaxy pairs \citep{Borthakur2015}. 
The strength of \Lyas absorption was found to sharply drop at $\rho \approx 136$~kpc, even when normalized by the target galaxy's $R_\mathrm{vir}$.
A similar decline has been seen by other surveys (e.g., \citealt{Keeney2017,Wilde2021,Klimenko2023}) and in simulations (e.g., \citealt{vandeVoort2019,Nelson2020}).
Stronger \Lyas absorption was also found to be positively correlated with larger \HIs mass in the galaxy's interstellar medium (ISM).
No correlation was observed, however, between \Lyas absorption and the galaxy's stellar mass or orientation with respect to its optical major axis.
These imply that \Lyas absorbers are able to trace the CGM structure around galaxies.

The combined COS-Halos~$+$~GASS sample found almost no \SiIIIs beyond $\sim$0.8~$R_\mathrm{vir}$ \citep{Borthakur2016}.
A positive correlation was seen, though, between the galaxy's star formation rate (SFR) and rest-frame equivalent width ($W_r$) of both \Lyas and \SiIII.
The kinematic properties of the sample showed no dependence on the galaxy's halo mass; thus, gravity alone cannot explain the observed dynamics.
These further indicated that \SiIIIs and \Lyas may be in the same physical regions, as was shown by \citet{Richter2016}.

Motivated by the discovery of COS-GASS that a galaxy's \HIs mass and the strength of $W_r$(\Lya) in its CGM are correlated \citep{Borthakur2015,Borthakur2016}, we have designed the Deciphering the Interplay between the ISM, Stars, and the CGM (DIISC) survey to trace the entire baryon cycle in the local Universe \citep{Gim2021,Padave2021,Padave2023,Padave2024,Borthakur2024}.
We add to the survey by investigating the CGM absorption features in the sample, which we refer to as COS-DIISC.
Here, we focus on the metal absorbers, while the \Lyas features, as well as a complete discussion of the survey design and goals, are discussed in \citet{Borthakur2024}.

There have been many absorption line spectroscopic surveys which have characterized the CGM across a range of galaxy properties and redshifts (e.g., KBSS, \citealt{Rudie2012}; COS-Halos, \citealt{Tumlinson2013}; COS-Dwarfs, \citealt{Bordoloi2014}; COS-GASS, \citealt{Borthakur2015}; COS-Burst, \citealt{Heckman2017}; COS-Weak, \citealt{Muzahid2018}; CASBaH, \citealt{Burchett2019}; CUBS, \citealt{Chen2020}; Project AMIGA, \citealt{Lehner2020}; CGM$^2$, \citealt{Wilde2021}).
The COS-DIISC program continues this by investigating the kinematics, content, and ionization state of the gas at the disk-CGM interface of galaxies as never before.

The rest of the paper is outlined as follows:
Section~\ref{sect:data} details the DIISC sample, the observations used in our analysis, and how we identified lines in our spectra.
Our analysis is described in Section~\ref{sect:results}.
Finally, Section~\ref{sect:summary} summarizes our results and main conclusions.
Throughout this paper, we assume a cosmology with $H_0 = 70$ \kmss Mpc$^{-1}$, $\Omega_m = 0.3$, and $\Omega_\Lambda = 0.7$.

\section{Sample \& Data} \label{sect:data}

\subsection{The DIISC Sample} \label{sect:sample}

The DIISC survey selected all galaxies with an FUV bright QSO (GALEX FUV magnitude brighter than 19) within $\sim$3.5~times their \HIs radii ($R_\mathrm{HI}$),\footnote{The \HIs radius of a galaxy is defined as the mean radius with a mass surface density $\Sigma$($M_{\mathrm{HI}}$) $= 1$~\Ms~pc$^{-2}$.} as calculated from the ALFALFA survey \citep{Giovanelli2005,Haynes2018} and the HIPASS survey \citep{Meyer2004,Zwaan2004}.
This led to a complete sample of 32 QSOs probing 34 local foreground galaxies.
For four galaxies, further observations revealed smaller \HIs radii than were found by the ALFALFA and HIPASS surveys, meaning their background QSOs probe the galaxies between 3.5 and 4.5~$R_\mathrm{HI}$.

We exclude two sight lines from the final sample. 
J0235-0925 because the data quality of the UV spectrum was too poor for fits to be performed.
In addition, J1024+2422 was excluded because of the presence of a Lyman limit system at a redshift of $z \approx 0.525$, leaving no flux below $\sim$1391 \AA.
This makes \SiIVs the only available species, which we did not detect.
We also exclude the galaxy KUG~1429+101 (probed by QSO J1432+0955) since it is at the same $z$ as NGC~5669, but with less mass and smaller radii; thus, we attribute the observed absorption to the larger galaxy.
This leaves 30 QSOs probing 31 foreground galaxies in the final sample.

\begin{figure}
    \centering
    \includegraphics[width=\linewidth]{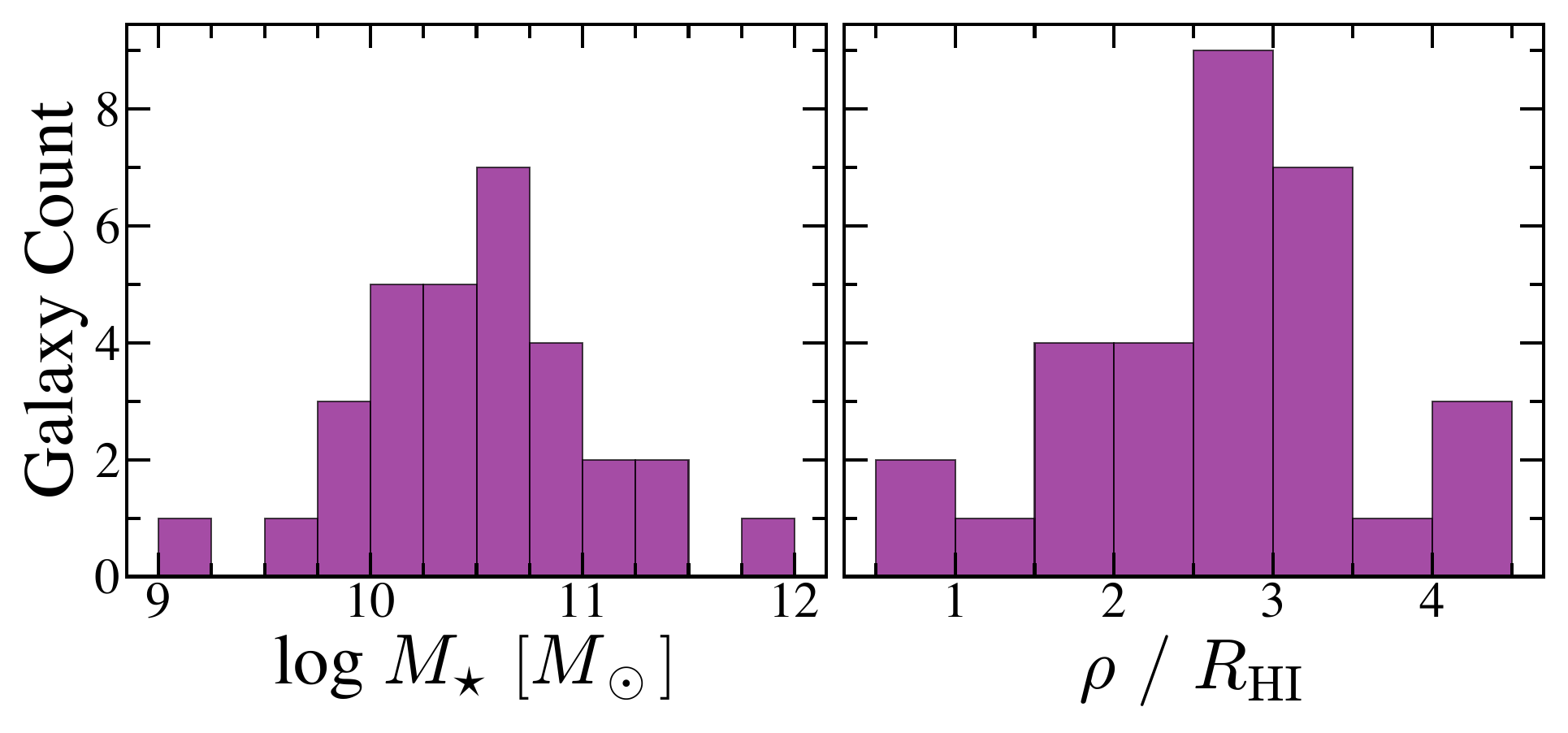}
    \caption{Distribution of stellar mass (left) and impact parameter relative to the \HIs radius of the targeted foreground galaxies (right) in the COS-DIISC sample.}
    \label{fig:diisc_hist}
\end{figure}

The left panel of Figure~\ref{fig:diisc_hist} shows the distribution of stellar mass ($M_\star$) of our sample, while the right panel indicates the impact parameter of the background QSOs normalized by the radius of the galaxy's $R_\mathrm{HI}$.
The DIISC galaxies cover a range of $M_\star$, log($M_\star$/\M) $= 8.5-12.0$, and have specific star formation rates (sSFR~$\equiv$~SFR~/~$M_\star$) between $10^{-12.05}-10^{-9.96}$~yr$^{-1}$.
Similarly, we cover a large range of impact parameters, $\rho = 21-163$~kpc. 
The properties of the COS-DIISC sample are listed in Table~\ref{tab:sample}.

\startlongtable
\begin{deluxetable*}{ccccccccccccC}
    \label{tab:sample}
    \tablecolumns{13}
    \setlength{\tabcolsep}{3.5pt}
    \tablecaption{Properties of the DIISC sample}
    \tablehead{
    \colhead{QSO} & \colhead{R.A.} & \colhead{Decl.} & \colhead{$z_\mathrm{QSO}$} & \colhead{Galaxy} & \colhead{$z_\mathrm{sys}$} & \colhead{$\rho$} & \colhead{$R_\mathrm{vir}$} & \colhead{$M_\mathrm{halo}$} & \colhead{$R_\mathrm{HI}$} & \colhead{$M_\mathrm{HI}$} & \colhead{$M_\star$} & \colhead{sSFR} \\
\colhead{(1)} & \colhead{(2)} & \colhead{(3)} & \colhead{(4)} & \colhead{(5)} & \colhead{(6)} & \colhead{(7)} & \colhead{(8)} & \colhead{(9)} & \colhead{(10)} & \colhead{(11)} & \colhead{(12)} & \colhead{(13)}}
    \startdata
    J0023+1547 & 5.878 & 15.796 & 0.4112 & NGC~99 & 0.0177 & 163 & 207 & 11.7 & 52.0 & 10.4 & 10.6 & -10.23 \\
    J0832+2431 & 128.084 & 24.517 & 1.2996 & KUG~0829+246B & 0.0432 & 41 & 173 & 11.5 & 33.4 & 10.0 & 10.4 & -10.30 \\
    J0835+2459 & 128.899 & 24.995 & 0.3299 & NGC~2611 & 0.0175 & 55 & 207 & 12.0 & 20.2 & 9.6 & 10.6 & -9.96 \\
    J0917+2719 & 139.369 & 27.331 & 0.0757 & J0917+2720 & 0.0469 & 52 & 203 & 12.0 & 26.8 & 9.8 & 10.6 & -10.88 \\
    J1042+2501 & 160.672 & 25.023 & 0.3426 & NGC~3344 & 0.0020 & 31 & 165 & 11.4 & 16.8 & 9.4 & 10.2 & -10.52 \\
    J1043+1151 & 160.900 & 11.858 & 0.7925 & M95/NGC~3351 & 0.0026 & 29 & 334 & 12.3 & 9.7 & 9.0 & 10.9 & -11.00 \\
    J1052+1017 & 163.086 & 10.298 & 0.2462 & NGC~3433 & 0.0091 & 128 & 252 & 12.0 & 41.0 & 10.2 & 10.8 & -10.77 \\
    J1059+1441 & 164.938 & 14.695 & 0.6306 & NGC~3485 & 0.0048 & 83 & 188 & 11.6 & 24.7 & 9.8 & 10.4 & -10.53 \\
    J1059+2517 & 164.995 & 25.286 & 0.6620 & 2MASXJ1059+2516 & 0.0207 & 38 & 145 & 11.3 & 14.6 & 9.3 & 9.9 & -10.09 \\
    J1124+1131 & 171.165 & 11.521 & 0.1429 & NGC~3666 & 0.0035 & 56 & 174 & 11.8 & 16.9 & 9.4 & 10.3 & -10.80 \\
    J1133+2423 & 173.358 & 24.391 & 0.5347 & NGC~3728 & 0.0232 & 120 & 382 & 12.8 & 68.8 & 10.6 & 11.3 & -10.84 \\
    J1134+2555 & 173.740 & 25.925 & 0.7098 & AGC~723761 & 0.0321 & 51 & 150 & 11.3 & 20.5 & 9.6 & 10.0 & -10.99 \\
    J1140+1136 & 175.192 & 11.614 & 0.6871 & NGC~3810 & 0.0033 & 41 & 187 & 11.6 & 16.7 & 9.4 & 10.4 & -10.34 \\
    J1144+0705 & 176.145 & 7.088 & 0.0749 & J1144+0703 & 0.0507 & 125 & 385 & 12.3 & 47.9 & 10.3 & 11.1 & -11.06 \\
    J1157+0906 & 179.291 & 9.102 & 0.3045 & AGC~213643 & 0.0367 & 88 & 139 & 11.2 & 33.3 & 10.0 & 9.8 & -10.60 \\
    J1157+1140 & 179.344 & 11.678 & 0.2905 & 2MASXJ1157+1139 & 0.0213 & 45 & 146 & 11.6 & 19.0 & 9.5 & 9.9 & -10.16 \\
    J1209+2616 & 182.322 & 26.270 & 0.5770 & UGC~7138 & 0.0072 & 37 & 119 & 11.0 & 11.2 & 9.1 & 9.2 & -10.14 \\
    J1221+0430 & 185.408 & 4.507 & 0.0947 & M61/NGC~4303 & 0.0052 & 26 & 457 & 12.7 & 28.9 & 9.9 & 11.4 & -10.67 \\
    J1223+1545 & 185.879 & 15.752 & 0.0806 & M100/NGC~4321 & 0.0053 & 39 & 204 & 11.7 & 15.6 & 9.4 & 10.6 & -10.18 \\
    J1231+1203 & 187.805 & 12.052 & 0.1162 & IC~3440 & 0.0260 & 83 & 163 & 11.4 & 24.2 & 9.7 & 10.2 & -10.53 \\
    J1301+2751 & 195.367 & 27.852 & 0.2425 & NGC~4921 & 0.0182 & 53 & 1009 & 13.8 & 12.6 & 9.2 & 11.8 & -12.05 \\
    J1303+2633 & 195.941 & 26.554 & 0.4356 & UGC~8161 & 0.0223 & 99 & 240 & 12.2 & 30.4 & 9.9 & 10.8 & -10.83 \\
    J1314+2606 & 198.696 & 26.107 & 0.0716 & J1314+2605 & 0.0431 & 96 & 215 & 12.1 & 33.1 & 10.0 & 10.7 & -10.27 \\
    J1354+1441 & 208.609 & 14.698 & 0.2143 & KUG~1352+149 & 0.0418 & 84 & 213 & 11.8 & 47.1 & 10.3 & 10.7 & -10.54 \\
    J1415+0445 & 213.775 & 4.763 & 0.2413 & UGC~9120 & 0.0192 & 90 & 323 & 12.3 & 22.0 & 9.7 & 11.1 & -10.98 \\
    J1432+0955 & 218.070 & 9.922 & 0.7663 & NGC~5669 & 0.0046 & 29 & 130 & 11.1 & 13.0 & 9.2 & 9.5 & -10.06 \\
    J1524+0419 & 231.199 & 4.322 & 0.7104 & 2MASXJ1524+0421 & 0.0374 & 107 & 221 & 12.1 & 25.5 & 9.8 & 10.7 & -10.39 \\
    J1533+1501 & 233.310 & 15.017 & 0.0909 & LEDA~140527 & 0.0429 & 22 & 149 & 11.3 & 24.9 & 9.8 & 10.0 & -10.33 \\
    J1533+1501 & 233.310 & 15.017 & 0.0909 & NGC~5951 & 0.0059 & 55 & 158 & 11.4 & 18.7 & 9.5 & 10.1 & -10.74 \\
    J1558+1205 & 239.591 & 12.093 & 0.5745 & IC~1149 & 0.0156 & 67 & 239 & 11.9 & 23.3 & 9.7 & 10.8 & -10.75 \\
    J1558+1255 & 239.731 & 12.932 & 0.2879 & KUG~1556+130 & 0.0347 & 109 & 176 & 11.8 & 30.5 & 9.9 & 10.4 & -10.59 \\
    \enddata
    \tablenotetext{}{Column (1) contains the GALEX name of the background QSO. Columns (2) and (3) show the R.A. and Decl.~(J2000) of the QSO. Column (4) indicates the QSO's redshift. Columns (5) and (6) show the name of the target galaxy and associated redshift, respectively. Column (7) is the impact parameter of the sight line in units of kpc. Columns (8) and (9) have the galaxy's virial radius in kpc and its halo mass in units of log~\M, respectively. The halo mass was calculated using Equation (21) from \citet{Behroozi2010}. Columns (10) and (11) indicate the galaxy's \HIs radius and mass in units of kpc and log~\M, respectively. The \HIs mass was measured by the ALFALFA survey \citep{Giovanelli2005,Haynes2018} with the radius being derived from the \HIs mass to radius relation found by \citet{Swaters2002}. Column (12) is the galaxy’s stellar mass in units of log~\M. Column (13) has the specific star formation rate (SFR~/~$M_\star$) in units of log~yr$^{-1}$ \citep{Padave2023}.}
\end{deluxetable*}

\subsection{Observations \& Line Identification} \label{sect:obs}

The observations used in our analysis were taken with the COS \citep{Osterman2011,Green2012} on board the Hubble Space Telescope (HST) through proposal GO-14071 (PI:~S.~Borthakur).
We used the medium-resolution grating G130M ($R$~$\approx$~15,000; FWHM~$\approx$~20~\kms) which has a wavelength coverage of $1140-1470$~\AA.
Our galaxies reside at $z \lesssim$ 0.05, giving us access to FUV transitions probing a range of gas phases. 
These include \Lyas ($\lambda1215$), \OIs ($\lambda1302$), \SiIIs ($\lambda\lambda1190, 1193, 1260, 1304$), \SiIIIs ($\lambda1206$), \SiIVs ($\lambda\lambda1393, 1402$), \CIIs ($\lambda1334$), and \NVs ($\lambda\lambda1238, 1242$).
The densities and temperatures traced by these transitions can be found in Figure~6 of \citet{Tumlinson2017}.

We used the standard COS pipeline to reduce our data set \citep{Rafelski2018}.
Data-analysis measurements were made using the same method employed by the COS-Halos \citep{Tumlinson2013,Werk2013}, COS-Dwarfs \citep{Bordoloi2014}, and COS-GASS \citep{Borthakur2015,Borthakur2016} surveys.
Individual exposures were coadded, and the spectra was binned by 3 pixels.
The COS pipeline is known to produce oversampled spectra, so binning by 3 pixels results in Nyquist sampling.
The continuum of each sight line was normalized by fitting Legendre polynomials between orders 1 and 5.
We identified all absorption features with rest-frame equivalent widths ($W_r$) at least $3\sigma$ above the noise level and within $\pm600$~\kmss of the galaxy's systematic velocity ($v_\mathrm{sys}$).
The absorbers beyond $\pm600$~\kmss of $v_\mathrm{sys}$ are likely not gravitationally bound to the galaxy and so are excluded in our analysis.
Absorbers not associated with our target galaxies were also identified to evaluate any contamination.

\subsection{Voigt Profile Analysis}

Voigt profiles were fit to the absorbers in our sample to measure the velocity centroids ($v_\mathrm{obs}$), column density ($N$), and Doppler width ($b$) of each component.
Table~\ref{tab:summary} summarizes the number of detected components of each ion along the sight lines.
Our rest-frame equivalent width measurements ($W_r$) and the results of our Voigt profile fits are available in the online journal. 
For the transitions that are not detected at $3\sigma$ above the noise level, we measure $W_r$ within $\pm$50~\kmss of the galaxy's systematic velocity ($v_\mathrm{sys}$) and report the $3\sigma$ value of the error as an upper limit.
For the QSO-galaxy pair J1144+0705-J1144+0703, \ion{S}{2} 1250 and 1253 from the Milky Way were found at the expected wavelength of \SiIIs 1190 and 1193, respectively, of the foreground galaxy.
We modeled and removed the \ion{S}{2} lines before finding the upper limit of the \SiIIs transitions.
In some cases, there were intervening absorbers at $v_\mathrm{sys}$ that we could not model and remove from the spectra to get a more representative measurement of $W_r$.

\begin{deluxetable*}{ccccccccc}
\centering
\tablecolumns{9}
\label{tab:summary}
\tablecaption{Number of Detected Components Along Each Sight Line}
\tablehead{
\colhead{QSO} & \colhead{Galaxy} & \colhead{$z_\mathrm{sys}$} & \colhead{\ion{O}{1}} & \colhead{\ion{Si}{2}} & \colhead{\ion{C}{2}} & \colhead{\ion{Si}{3}} & \colhead{\ion{Si}{4}} & \colhead{\ion{N}{5}}  \\
\colhead{(1)} & \colhead{(2)} & \colhead{(3)} & \colhead{(4)} & \colhead{(5)} & \colhead{(6)} & \colhead{(7)} & \colhead{(8)} & \colhead{(9)}}
\startdata
J0023+1547 & NGC 99 & 0.0177 & 0 & 0 & 0 & 0 & 0 & 0 \\
J0832+2431 & KUG 0829+246B & 0.0430 & 2 & 2 & 1 & 2 & --- & 0 \\
J0835+2459 & NGC 2611 & 0.0175 & 3 & 2 & 3 & 4 & 7 & 0 \\
J0917+2719 & J0917+2720 & 0.0469 & 1 & 3 & 2 & 2 & --- & 0 \\
J1042+2501 & NGC 3344 & 0.0020 & 0 & 2 & 1 & 1 & 1 & 1 \\
J1043+1151 & M95/NGC 3351 & 0.0026 & 0 & 4 & 1 & 1 & 1 & 1 \\
J1052+1017 & NGC 3433 & 0.0091 & 0 & 0 & 0 & 0 & 0 & 0 \\
J1059+2517 & 2MASXJ1059+2516 & 0.0207 & 0 & 0 & 0 & 0 & 0 & 0 \\
J1059+1441 & NGC 3485 & 0.0048 & 0 & 0 & 0 & 0 & 0 & 0 \\
J1124+1131 & NGC 3666 & 0.0035 & 0 & 1 & 2 & 2 & 2 & 0 \\
J1133+2423 & NGC 3728 & 0.0232 & 0 & 4 & 3 & 3 & 0 & 0 \\
J1134+2555 & AGC 723761 & 0.0321 & 0 & 0 & 0 & 0 & 0 & 0 \\
J1140+1136 & NGC 3810 & 0.0033 & 0 & 2 & 1 & 3 & 2 & 1 \\
J1144+0705 & J1144+0703 & 0.0507 & 0 & 0 & 0 & 2 & --- & --- \\
J1157+0906 & AGC 213643 & 0.0367 & 0 & 0 & 1 & 1 & 2 & 0 \\
J1157+1140 & 2MASXJ1157+1139 & 0.0213 & 0 & 0 & 1 & 1 & 1 & 0 \\
J1209+2616 & UGC 7138 & 0.0072 & 0 & 0 & 1 & 0 & 0 & 0 \\
J1221+0430 & M61/NGC 4303 & 0.0052 & 2 & 6 & 5 & 6 & 7 & 3 \\
J1223+1545 & M100/NGC 4321 & 0.0053 & 0 & 0 & 0 & 0 & 0 & 0 \\
J1231+1203 & IC 3440 & 0.0260 & 0 & 0 & 0 & 0 & 0 & 0 \\
J1301+2751 & NGC 4921 & 0.0182 & 0 & 0 & 0 & 0 & 0 & 0 \\
J1303+2633 & UGC 8161 & 0.0223 & 0 & 0 & 0 & 0 & 0 & 0 \\
J1314+2606 & J1314+2605 & 0.0431 & 0 & 1 & 0 & 2 & --- & 0 \\
J1354+1441 & KUG 1352+149 & 0.0418 & 0 & 0 & 0 & 4 & --- & 0 \\
J1415+0445 & UGC 9120 & 0.0192 & 0 & 0 & 0 & 0 & 0 & 0 \\
J1432+0955 & NGC 5669 & 0.0046 & 2 & 2 & 2 & --- & 3 & 0 \\
J1524+0419 & 2MASXJ1524+0421 & 0.0374 & 0 & 1 & 2 & 2 & 0 & 0 \\
J1533+1501 & LEDA 140527 & 0.0429 & 1 & 2 & 2 & 3 & --- & 0 \\
J1533+1501 & NGC 5951 & 0.0059 & 0 & 0 & 0 & 0 & 0 & 0 \\
J1558+1205 & IC 1149 & 0.0156 & 1 & 1 & 1 & 1 & 1 & 0 \\
J1558+1255 & KUG 1556+130 & 0.0347 & 0 & 0 & 0 & 0 & --- & --- \\
\enddata
\tablenotetext{}{Column (1) shows the QSO of the sight line. Columns (2) and (3) are the foreground galaxy and its redshift. Columns (4)-(9) indicates the number of components of each ion detected at 3$\sigma$ above the noise level. Ions that not covered by our data are indicated by dashes.}
\end{deluxetable*}

\begin{figure*}
    \centering
    \includegraphics[width=0.9\linewidth]{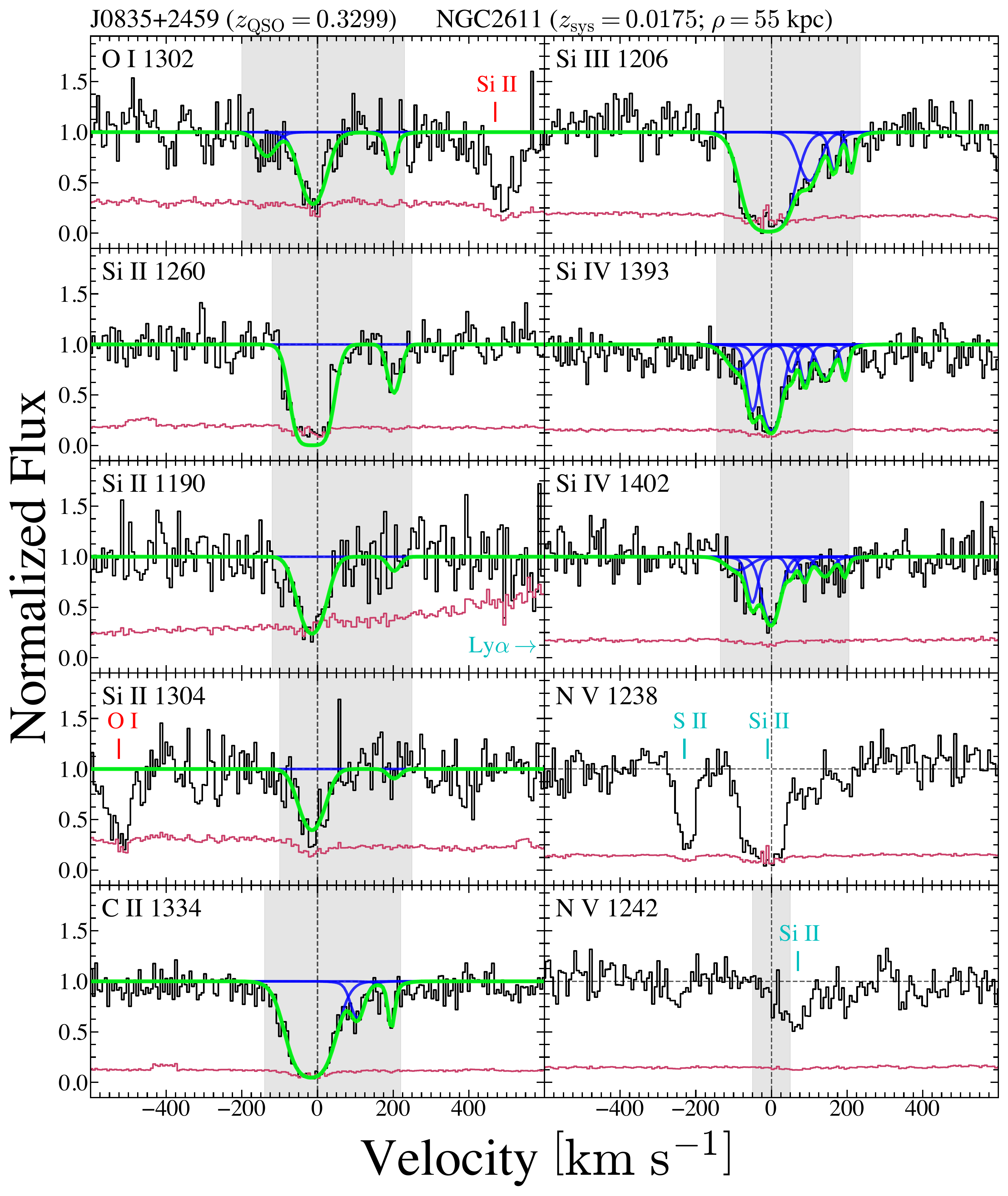}
    \caption{Voigt profile fits of the spectrum towards the QSO J0835+2459 probing the galaxy NGC~2611. Each panel is centered on the galaxy's systematic velocity ($v_\mathrm{sys}$). Individual components are shown in blue, with the combined profile in green. The gray shaded regions show the range used to calculate the rest-frame equivalent width. Features other than the one being highlighted are colored based on their origin: cyan for Milky Way features, red for features associated with our target galaxy, yellow for those from the background QSO, and pink for intervening systems. The \SiIIs $\lambda$1193 \AA\ transition is located within the \Lyas emission line and so is not included in the fit or shown here. Similar figures are available for each QSO-galaxy pair in the online journal.}
    \label{fig:ex_slice}
\end{figure*}

We present an example of our analysis in Figure~\ref{fig:ex_slice} for the QSO-galaxy pair J0835+2459-NGC~2611.
The species being plotted as well as its rest wavelength, in units of \AA, are shown in the upper left of each panel.
The normalized flux and uncertainties, centered on the galaxy's $v_\mathrm{sys}$, are shown in black and dark pink, respectively.
The regions used to calculate $W_r$ are shown in gray.
Individual Voigt profile components are shown in blue, while the combined profile is in green.
Any additional features within $\pm$600~\kmss of the galaxy's $v_\mathrm{sys}$ are labeled as follows:
Milky Way features are shown in cyan, those associated with our target galaxy are in red, those from the background QSO are yellow, while any other lines (i.e., ``intervening'' systems) are pink.

The QSO-galaxy pair J1042+2501-NGC~3344 was previously fit by \citet{Padave2021} to study the galaxy's extended UV disk.
We report their measurements for \HI, \SiII, \SiIII, \SiIV, and \CII, rather than fitting the absorption features again due to the complex nature of the fit and low signal-to-noise values.
The Voigt profile fits for this pair are shown in \citet{Padave2021}.
Most metals had a single component, while the \SiIIs absorbers had two components.
The low-$z$ nature of NGC~3344 caused the blueward component of its \Lyas feature to be blended with the Milky Way's \Lyas line.
However, blending was not an issue for any of the metal features.

\section{Results \& Discussion} \label{sect:results}

\subsection{Absorber Kinematics} \label{sect:kin}

\begin{figure}
    \centering
    \includegraphics[width=\linewidth]{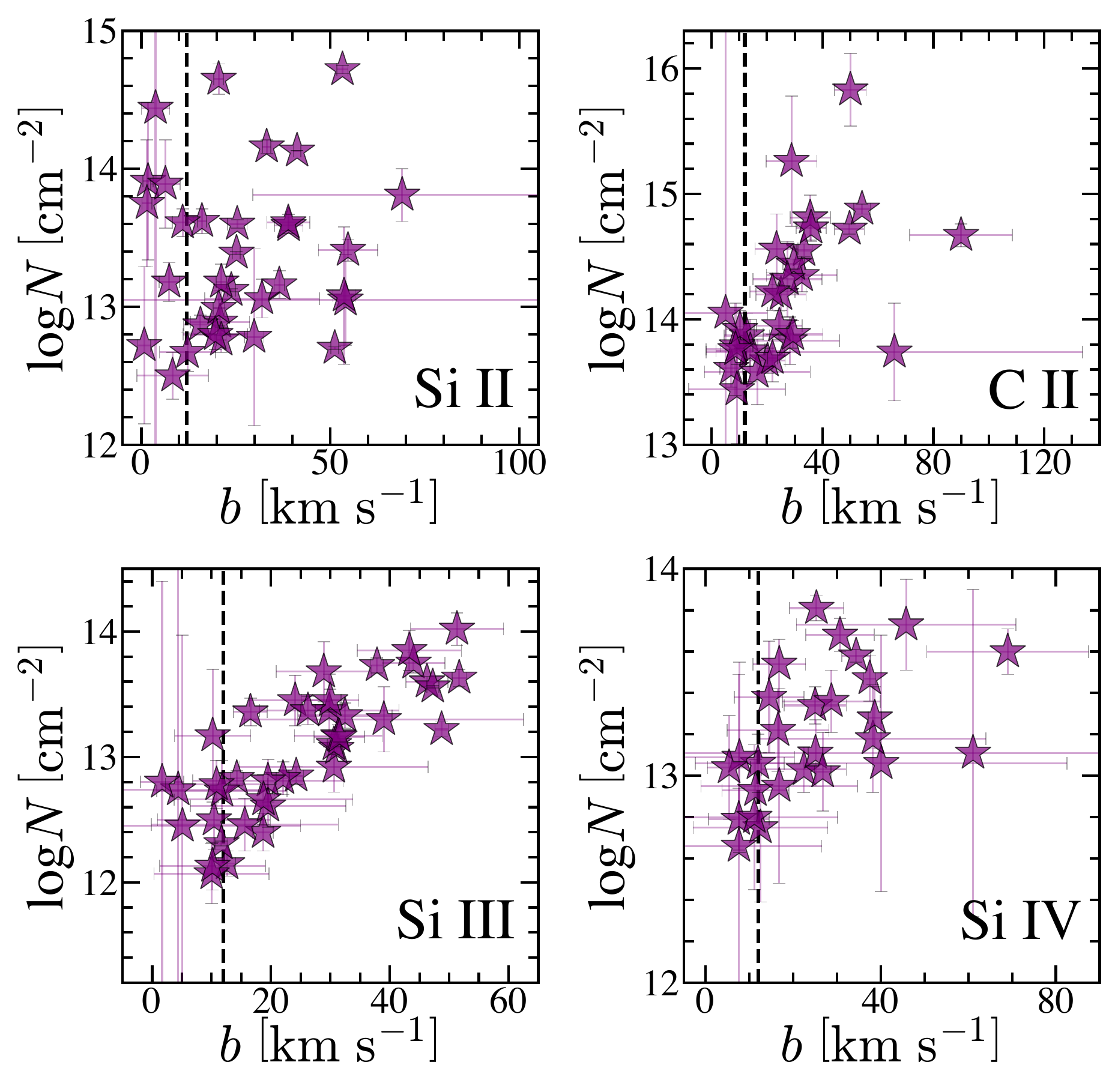}
    \caption{Measured Doppler width against column density for our most frequently detected ions. The ion being shown in each panel is indicated in the lower right corner. The vertical black dashed line indicates the resolution of the G130M filter of COS (FWHM $\approx$ 20~\kmss corresponding to a $b \approx 12$~\kms). Components to the left of the line are likely unresolved. The strength of the correlation seems to be related to the number of available transitions. It is strongest for \SiIIIs and \CIIs which have a single transitions; weak for \SiIV, which has two transitions; and not present for \SiIIs which has four transitions. \CIIs and \SiIIIs are also more likely to be saturated than \SiIIs and \SiIV, which could also playing a role in causing this trend.} 
    \label{fig:logN_b}
\end{figure}

We looked for trends between the Doppler width of absorbers ($b$) with various galaxy properties in Table~\ref{tab:sample}.
A correlation between $b$ and column density (log$N$) was observed in each ion except \SiII, as shown in Figure~\ref{fig:logN_b}.
Smaller log$N$ values seem to give rise to narrower absorption features, while larger log$N$ values generally have wider features.
Performing Kendall~$\uptau$ tests revealed that the trends in \CII, \SiIII, and \SiIVs are statistically significant with $\uptau$ values of 0.49, 0.61, and 0.43, with associated $p$-values of $\sim$10$^{-4}$, $\sim$10$^{-8}$, and $\sim$10$^{-3}$, respectively.
\SiII, on the other hand, is consistent with no correlation ($\uptau = 0.17$ and $p$-value = 0.18).

One explanation for this trend is that \CIIs and \SiIIIs are more likely to be saturated than \SiIIs and \SiIV.
Another explanation is the presence of unresolved, overlapping components appearing as a single component along sight lines probing closer to the galactic disk, similar to what has been seen in \MgIIs (e.g., \citealt{Kacprzak2011,Nielsen2018}).
Of the 15 sight lines that showed \SiIIs $\lambda$1260 absorption, 12 had simultaneous detections in at least one of the weaker transitions ($\lambda\lambda1193, 1190, 1304$).
This prevented the trend from emerging in the ion since the weaker lines allow us to resolve more components when the $\lambda1260$ line is saturated.
However, for the \SiIVs ($\lambda\lambda1393,1402$) and \NVs ($\lambda\lambda1238,1242$) doublets, only the stronger transition was detected in 7 of 11 and 2 of 4 sight lines, respectively.
Not having the weaker counterpart may have caused the trend to be present in these ions, though not as strongly as \CIIs and \SiIII.

The largest (primary) components of \SiII, \CII, and \SiIIIs along each sight line, as measured by log$N$, tend to have velocity centroids ($v_\mathrm{obs}$) within $\pm$100~\kmss of the galaxy's $v_\mathrm{sys}$.
Meanwhile, the smaller (secondary) components are found as far away as $-300$ and 200~\kms.
At distances farther from the galaxies (i.e., larger values of $\rho$) these ions produced few absorption features, implying that the cool CGM is more dense closer to galaxies than farther into the halo.
We also find ions over a large range of ionization potentials (IPs), e.g., 13.6 eV for \OIs and 97.9 eV for \NV, at the similar velocities, indicating kinematic alignment across multiple gas phases.
However, it is possible that absorption from multiple phases that are kinematically aligned may not arise from the same process \citep{Cooper2021,Haislmaier2021,Qu2022}.

\begin{figure*}
    \centering
    \includegraphics[width=0.49\linewidth]{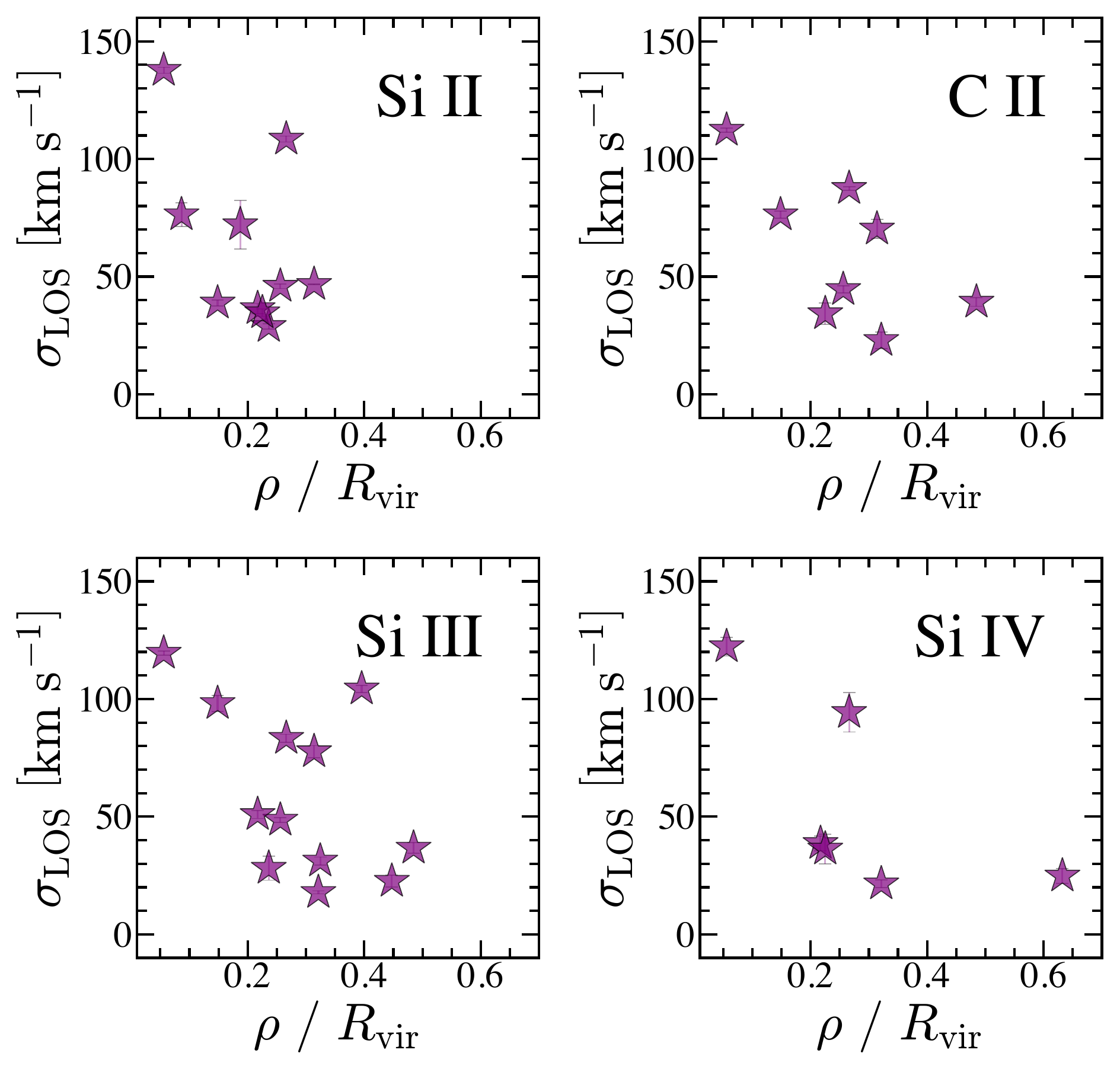}
    \includegraphics[width=0.49\linewidth]{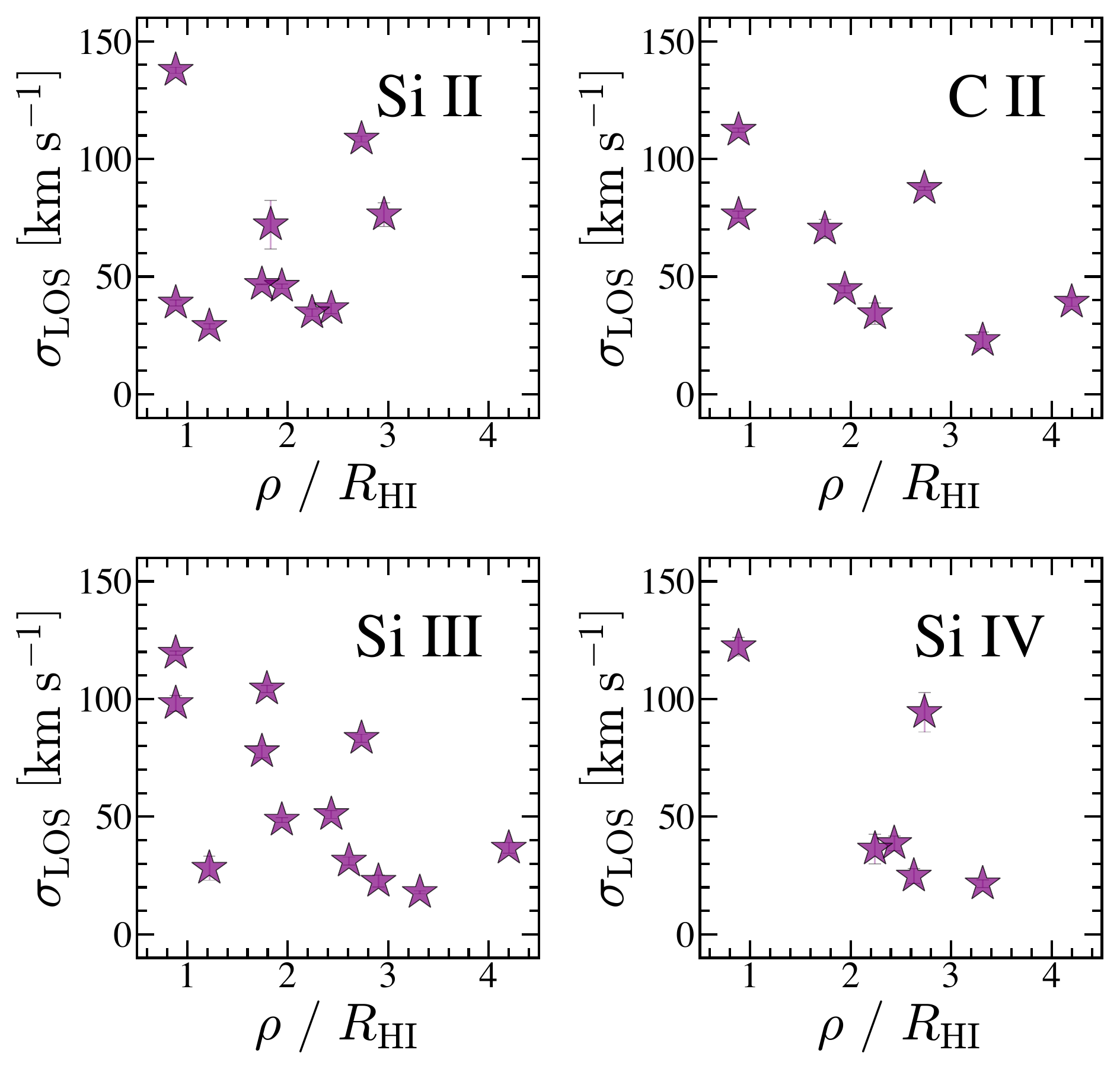}
    \caption{Inferred line-of-sight velocity dispersion as a function of impact parameter relative to the galaxy's virial radius (left) and the galaxy's \HIs radius (right) for the most frequently detected ions. The ion being shown in each panel is indicated in the upper right corner.}
    \label{fig:turb}
\end{figure*}

The amount of motion along a sight line can be inferred by measuring the line-of-sight velocity dispersion. 
We define this measure as $\sigma_{\mathrm{LOS}}^{2} \equiv \langle \left[v - \langle v\rangle \right]^2 \rangle$, where $v$ represents the velocity centroid of the components and the angled brackets indicate taking the mean.
Sight lines with a single component are excluded from this analysis.
This measure has been shown to trace the underlying turbulent velocities in the CGM on average \citep{Koplitz2023}.
For our most frequently detected ions (\SiII, \CII, \SiIII, and \SiIV), we show in Figure~\ref{fig:turb} the \sigls values as a function of $\rho$ normalized by the galaxy's $R_\mathrm{vir}$ (left panel) and by $R_\mathrm{HI}$ (right panel).
Most of the panels in the figure show anticorrelations, except for \SiIIs when $\rho$ is normalized by the galaxy's $R_\mathrm{HI}$.
Kendall~$\uptau$ tests indicate that the correlations of \SiII, \SiIII, and \SiIVs are not statistically significant.
With \CII, however, these tests show that the correlation is stronger when $\rho$ is normalized by $R_\mathrm{HI}$, with $\uptau = -0.48$ ($p$-value = 0.031) compared to $\uptau = -0.33$ ($p$-value = 0.153) for $\rho$ / $R_\mathrm{vir}$, respectively.
The \sigls values we measure, as well as the radial decline out to $\sim$0.4~$R_\mathrm{vir}$, are similar to those derived from the \SiIIs measurements of the COS-Halos survey \citep{Werk2013} and the MAIHEM simulations \citep{Koplitz2023}.

\begin{figure}
    \centering
    \includegraphics[width=\linewidth]{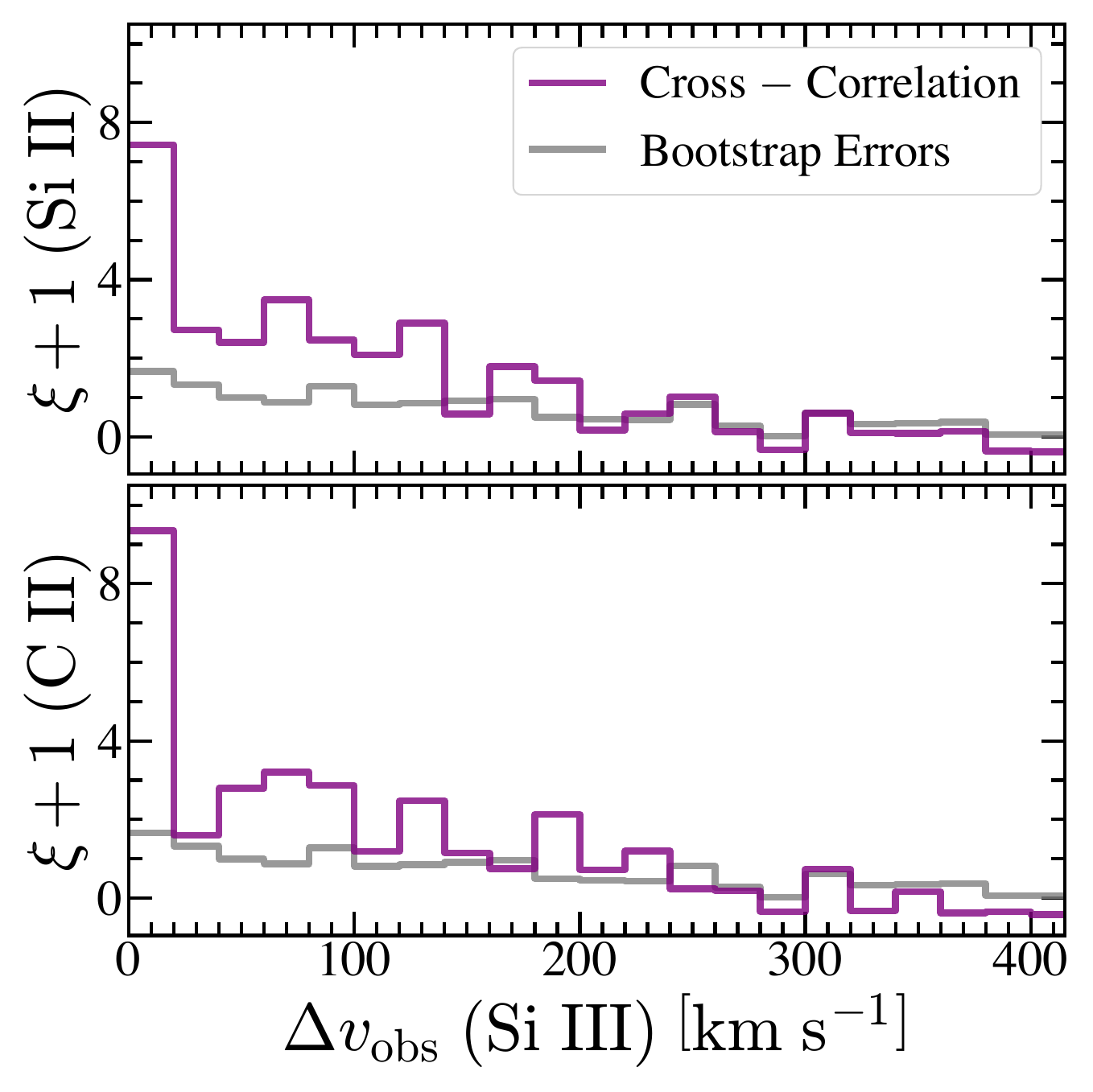}
    \caption{Cross-correlation analysis of \SiIIIs absorbers shown in purple with the standard deviation of our bootstrap analysis (as discussed in Section~\ref{sect:kin}) in gray. The top panel compares the velocity centroids of \SiIIs to those of \SiIIIs while the bottom panel compares the centroids of \CIIs to \SiIII.}
    \label{fig:bootstrap}
\end{figure}

The velocity centroids of our Voigt profile fits can also be used to  determine how clustered the components are in velocity space.
To constrain this, we performed a cross-correlation analysis between $v_\mathrm{obs}$ of \SiIIs and \CIIs against \SiIII.
We compare to \SiIIIs since it is the most frequently detected ion in our sample.
The generalized \citet{LandySzalay93} estimator was adopted such that 

\begin{equation}
    \xi + 1 = \frac{(\mathrm{DD} / n_\mathrm{DD}) - 2(\mathrm{DR} / n_\mathrm{DR}) + (\mathrm{RR} / n_\mathrm{RR})}{(\mathrm{RR} / n_\mathrm{RR})}
\end{equation}

where DD refers to data-data matching, DR refers to data-random matching, and RR refers to random-random matching.
Following \citet{Tejos2014} and \citet{Finn2016}, each matching set was normalized, such that $n_\mathrm{DD} = N_\mathrm{real} (N_\mathrm{real} - 1) / 2$, $n_\mathrm{DR} = N_\mathrm{rand} (N_\mathrm{real})^{2}$, and $n_\mathrm{RR} = N_\mathrm{rand} N_\mathrm{real} (N_\mathrm{rand} N_\mathrm{real} - 1) / 2$.
Here, $N_\mathrm{real}$ and $N_\mathrm{rand}$ refer to the number of real and random absorbers, respectively used in the analysis.
The random samples were drawn from a uniform distribution between $\pm$600~\kms.
Our uncertainties were estimated using a bootstrap analysis in which we randomly select sight lines, with replacements, and performed the same cross-correlation analysis.
This was done 1000 times, with the 1$\sigma$ width of this distribution adapted as our uncertainties.
The results of this analysis are shown in Figure~\ref{fig:bootstrap}.

Comparing $v_\mathrm{obs}$(\SiII) to $v_\mathrm{obs}$(\SiIII) in the top panel of Figure~\ref{fig:bootstrap} shows a peak in the velocity separations at $\Delta v_\mathrm{obs} \leq 20$~\kmss with a decreasing trend out to $\sim$400~\kmss with no pair of absorbers being found with further separations.
There are three bins that are statistically significant at $\geq$3$\sigma$: $\leq$20~\kms, 60 $-$ 80~\kms, and 120 $-$ 140~\kms.
Performing this analysis between $v_\mathrm{obs}$(\SiIII) and $v_\mathrm{obs}$(\CII), as shown in the bottom panel of Figure~\ref{fig:bootstrap}, finds a similar peak at $\Delta v_\mathrm{obs} \leq 20$~\kms, with an even sharper falloff.
Similarly, no pair of absorbers were found with $\Delta v_\mathrm{obs} \gtrsim 400$~\kms.
Three bins in this distribution are found to be statistically significant: $\leq$20~\kms, 60 $-$ 80~\kms, and 180 $-$ 200~\kms.
It is important to note that the noise dominates this analysis at velocity separations $\geq$200~\kmss in both panels.
Performing Kolmogorov-Smirnov tests between these distributions returned a $p$-value near 1, suggesting that these samples are not different at a statistically significant level.

When considering individual QSO-galaxy pairs, we find that galaxies with larger $M_\mathrm{\star}$ and $M_\mathrm{halo}$ tend to have larger velocity separations.
This could be indicative of virial support of the CGM gas.

A similar cross-correlation analysis was done by \citet{Borthakur2016} between the \Lyas and \SiIIIs absorbers from the combined COS-Halos$+$COS-GASS sample.
Their distribution followed a similar trend as those shown in Figure~\ref{fig:bootstrap}, with a peak in their lowest $\Delta v_\mathrm{obs}$ bin and a decline out to $\sim$400~\kms.
However, they find a few systems with $\Delta v_\mathrm{obs} > 400$~\kms.
The consistency between these analyses suggest that \Lya, \SiII, \CII, and \SiIIIs may be cospatial or tracing similar gas phases.

\subsection{Metal Content \& Equivalent Widths} \label{sect:ew}

\begin{figure*}
    \centering
    \includegraphics[width=\linewidth]{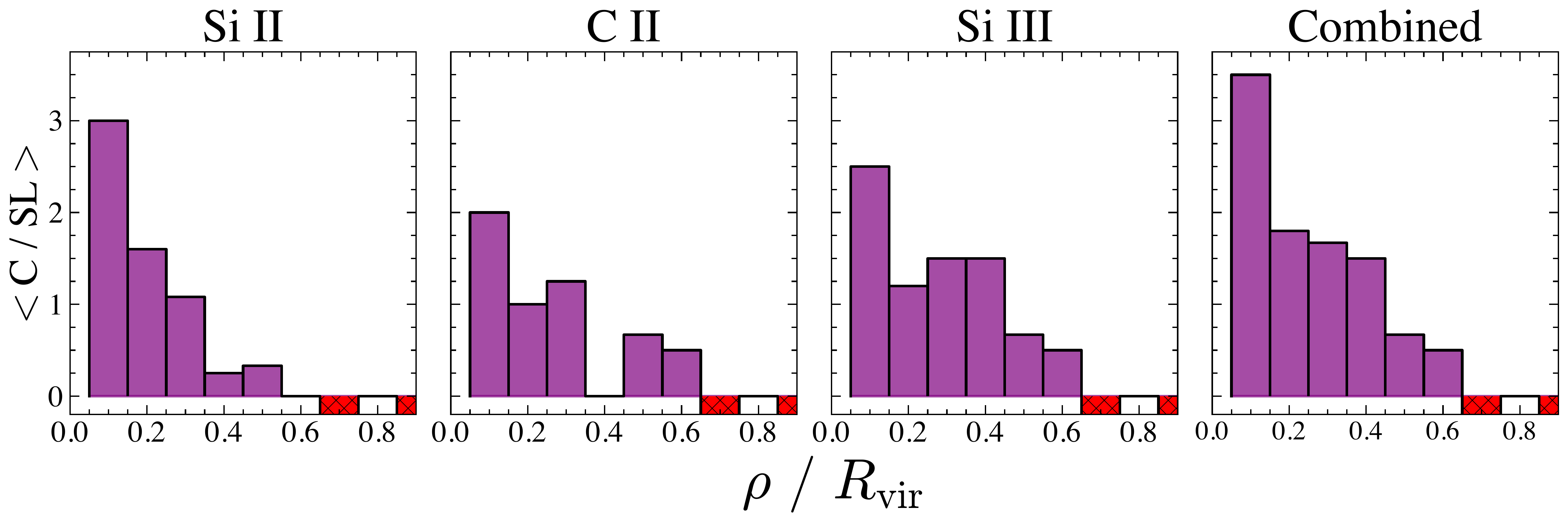}
    \caption{Histograms of the average number of components per sight line ($<$C/SL$>$) of \SiIIs (far left), \CIIs (left center), and \SiIIIs (right center) as a function of impact parameter normalized by the galaxy's virial radius. The far right panel is the number of components along a sight line is determined by considering the \SiII, \CII, and \SiIIIs features together, as defined in the text. Bins that do not contain any sight lines are indicated by red hatching to distinguish them from bins in which no components were detected.}
    \label{fig:ncomps}
\end{figure*}

We find that the sight lines that probe the CGM closer to the foreground galaxy, relative to both $R_\mathrm{vir}$ and $R_\mathrm{HI}$, have more components on average than those farther out.
We show this distribution for $\rho / R_\mathrm{vir}$ in Figure~\ref{fig:ncomps}.
By considering the number of components of each ion along a sight line, we are able limit the impact that unresolved components and weak absorbers have on the individual distributions.
We present the results of this in the far right panel of Figure~\ref{fig:ncomps}.
Similar to what was seen when considering each ion individually, there are more resolved components closer to the foreground galaxy.
These distributions are not consistent from the radial density distribution of a Navarro–Frenk–White dark matter halo \citep{Navarro1996} with $R_\mathrm{vir} = 213$~kpc and $M_\mathrm{vir} = 10^{11.8}$~\M, the median values of the COS-DIISC sample.
In addition, the distributions are not consistent with being produced by longer path lengths through a CGM of uniform density.
It is important to note that we are comparing the shape of these distributions to the average number of components per sight line, not predicting the number of absorbing clouds they would produce.

\begin{figure*}
    \centering
    \includegraphics[width=\linewidth]{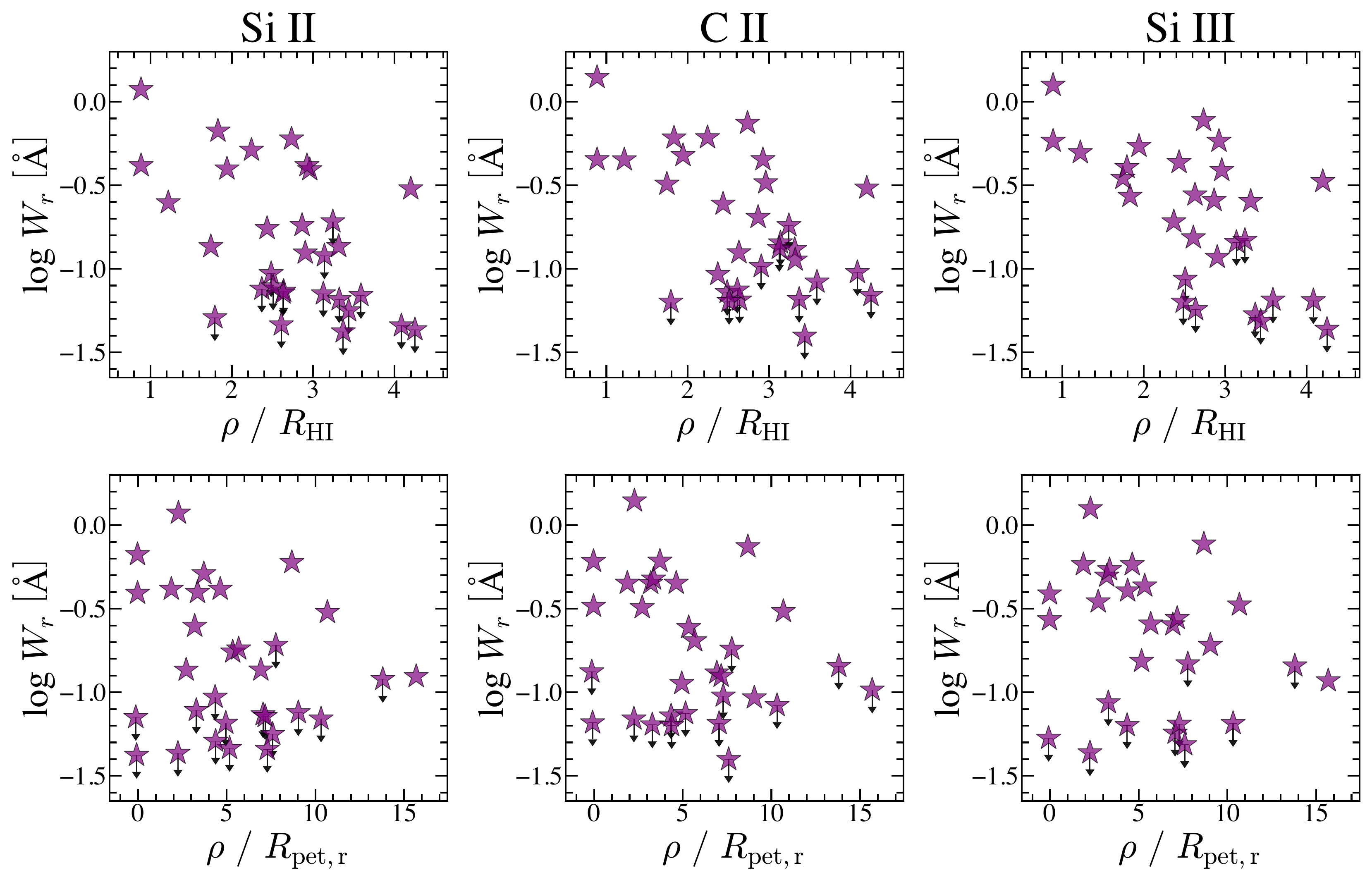}
    \caption{Measured rest-frame equivalent width of \SiIIs (left column), \CIIs (center column), and \SiIIIs (right column) as a function of impact parameter normalized by the galaxy's \HIs disk (top panels) and impact parameter normalized by the galaxy's $r$-band Petrosian radius (bottom panels). Upper limits are shown with a downward arrow.}
    \label{fig:Wr_rho_RHI_Rpetr}
\end{figure*}

Since the DIISC sample was selected based on the QSO's position relative to the galaxy's $R_\mathrm{HI}$, we are able to determine the distribution of absorbers in terms of distance from the galaxy's \HIs disk.
In the top panels Figure~\ref{fig:Wr_rho_RHI_Rpetr}, we show the distribution of log$W_r$ for \SiIIs (left columns), \CIIs (center columns), and \SiIIIs (right columns) as a function of $\rho$~/~$R_\mathrm{HI}$.
A clear anticorrelation emerges with sight lines closer to the \HIs disks tending to have larger log$W_r$ values than farther away.
It is noteworthy that all sight lines with \SiIIIs nondetections are found beyond $\sim$2.5~$R_\mathrm{HI}$.
\SiIIs and \CII, however, have a nondetection at $\rho \approx 1.8$~$R_\mathrm{HI}$ from the QSO-galaxy pair J1354+1441-KUG1352+149 where \SiIIIs is detected.
The \Lyas and \SiIIIs measurements from the COS-GASS survey also produced anticorrelations with distance from the \HIs disk out to $\sim$30~$R_\mathrm{HI}$ \citep{Borthakur2015}.
These show that the distance a QSO probes with respect to a galaxy's \HIs disk will largely dictate whether or not metals with low and medium IPs (13.6 eV $\leq$ IP $\leq$ 33.5 eV) are found along the sight line.

A similar radial decline was seen in the \HIs absorbers of MaNGA galaxies when looking at $\rho$ relative to the galaxy's effective radius of the stellar disk \citep{Klimenko2023}.
Our measurements of \SiII, \CII, and \SiIIIs display a similar trend with respect to the galaxy's $r$-band Petrosian radius ($R_\mathrm{pet,r}$).
We show these trends in the bottom panels of Figure~\ref{fig:Wr_rho_RHI_Rpetr}.
Unlike the relationships seen with respect to $R_\mathrm{HI}$, the nondetections are found across the entire range of $\rho~/~R_\mathrm{pet,r}$ for these ions.
As a result, we conclude that the distance from a galaxy's \HIs disk is a better tracer of metals in the CGM.

\begin{figure*}
    \centering
    \includegraphics[width=\textwidth]{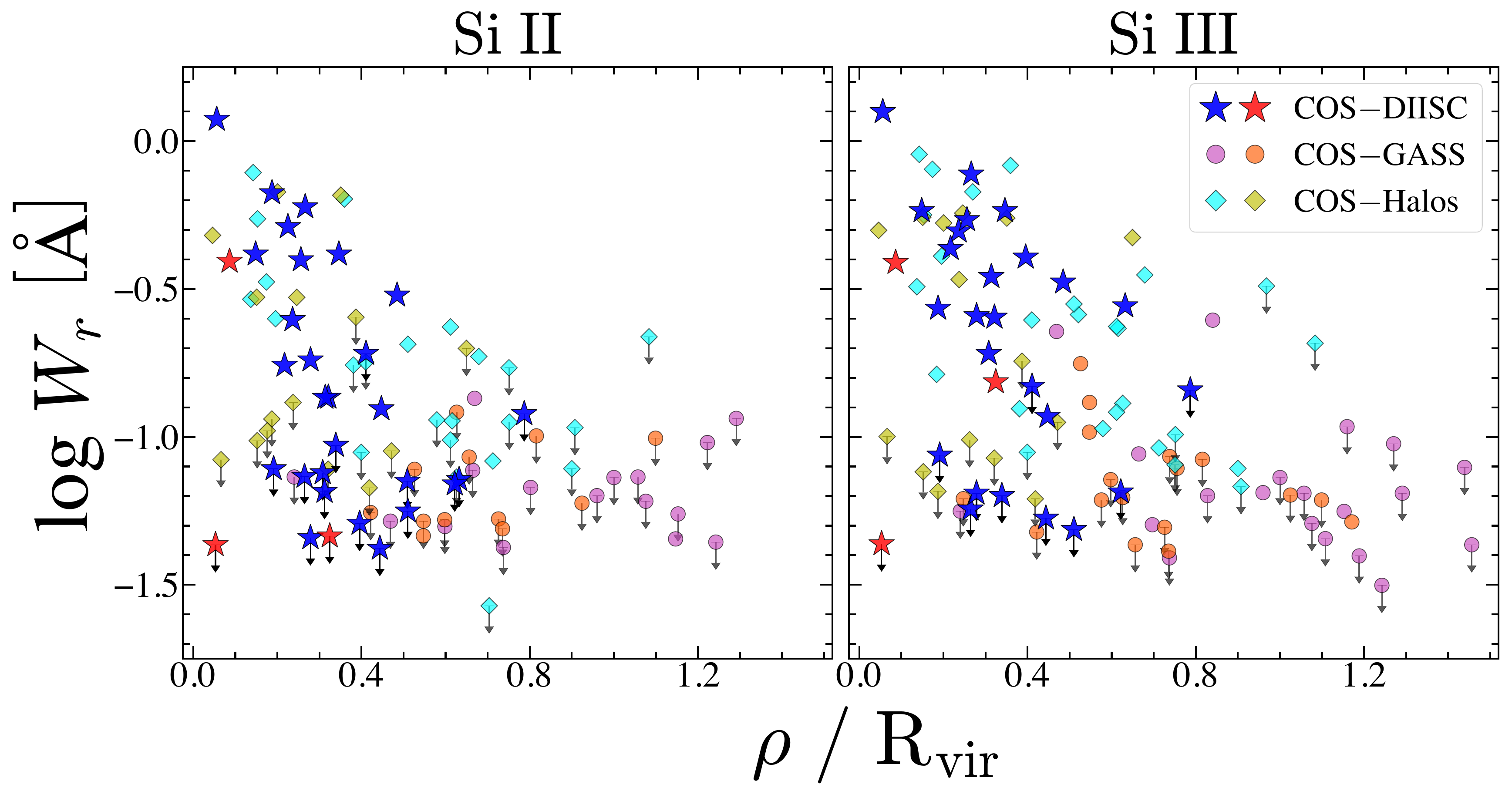}
    \caption{Measured rest-frame equivalent width of \SiIIs (left) and \SiIIIs (right) as a function of impact parameter relative to the galaxy's virial radius. We show the different samples being plotted as different symbols with stars representing COS-DIISC points, circles showing COS-GASS, and diamonds for COS-Halos. Each sample is split into star-forming (shown in blue for COS-DIISC, navy for COS-GASS, and cyan for COS-Halos) and passive (shown in red for COS-DIISC, orange for COS-GASS, and yellow for COS-Halos) galaxies as defined in Section~\ref{sect:ew}. Upper limits are plotted with a downward arrow.}
    \label{fig:Wr_rho_Rvir}
\end{figure*}

With the COS-Halos and COS-GASS surveys covering a similar range of galactic parameters as we do in COS-DIISC, we can combine the samples to study global trends in the CGM.
We look for trends between this combined sample with respect to $\rho$ normalized by the galaxy's $R_\mathrm{vir}$ since the $R_\mathrm{HI}$ of COS-Halos galaxies are unknown.
In Figure~\ref{fig:Wr_rho_Rvir}, we show log$W_r$ of \SiIIs (left) and \SiIIIs (right) as a function of $\rho$~/~$R_\mathrm{vir}$.
The foreground galaxies in each sample are separated into star-forming and passive galaxies based on their specific star formation rates (sSFR $\equiv$ SFR~/~$M_\star$) being above or below $10^{-11}$ yr$^{-1}$, respectively.
The star-forming galaxies are shown as blue stars for COS-DIISC, navy circles for COS-GASS, and cyan diamonds for COS-Halos, while passive galaxies are shown as red stars for COS-DIISC, orange circles for COS-GASS, and yellow diamonds for COS-Halos.

Larger values of log$W_r$ (i.e., stronger absorbers) are found closer to galaxies, similar to what we see in Figure~\ref{fig:Wr_rho_RHI_Rpetr}.
Beyond $\sim$0.4~$R_\mathrm{vir}$, no absorbers were detected with log$W_r$(\SiII) $> -0.5$.
This suggests that the outer halos of these galaxies contain weaker absorbing clouds than are found closer to the disk though this could be the result of difference in the density or ionization state of the gas.
Figure~\ref{fig:Wr_rho_Rvir} shows that few metals are detected beyond $\sim$0.7~$R_\mathrm{vir}$, consistent with previous studies (e.g., \citealt{Liang2014,Lehner2015,Qu2023}).
A similar radial decline was seen in the total silicon column density ($N(\mathrm{Si}) = N(\mathrm{Si\ II}) + N(\mathrm{Si\ III}) + N(\mathrm{Si\ IV})$) for M31 \citep{Lehner2020}.
The vast majority of upper limits of both ions have log$W_r < -1$.
Similar trends have been found in \MgIIs (e.g., \citealt{Chen2010, Churchill2013, Nielsen2013a, Ho2017}).

Comparing Figure~\ref{fig:Wr_rho_Rvir} to the top panels of Figure~\ref{fig:Wr_rho_RHI_Rpetr} reveals an interesting trend. 
Nondetections (upper limits) are only found at $\rho \gtrsim 1.5$~$R_\mathrm{HI}$ in Figure~\ref{fig:Wr_rho_RHI_Rpetr} while they are found across the entire range of $\rho$~/~$R_\mathrm{vir}$ in Figure~\ref{fig:Wr_rho_Rvir}.
The \Lyas absorbers in the COS-DIISC sample produced a similar trend \citep{Borthakur2024}.
This likely indicates that the radial profile of absorbers in the CGM is better traced by $\rho$~/~$R_\mathrm{HI}$ rather than $\rho$~/~$R_\mathrm{vir}$, which is typically used in CGM studies.
In particular, this may show that the strength of absorbers in the inner CGM (radii $\lesssim$2.5 $R_\mathrm{HI}$) can largely be predicted by the size of the galaxy's \HIs disk.

\begin{figure*}
    \centering
    \includegraphics[width=\linewidth]{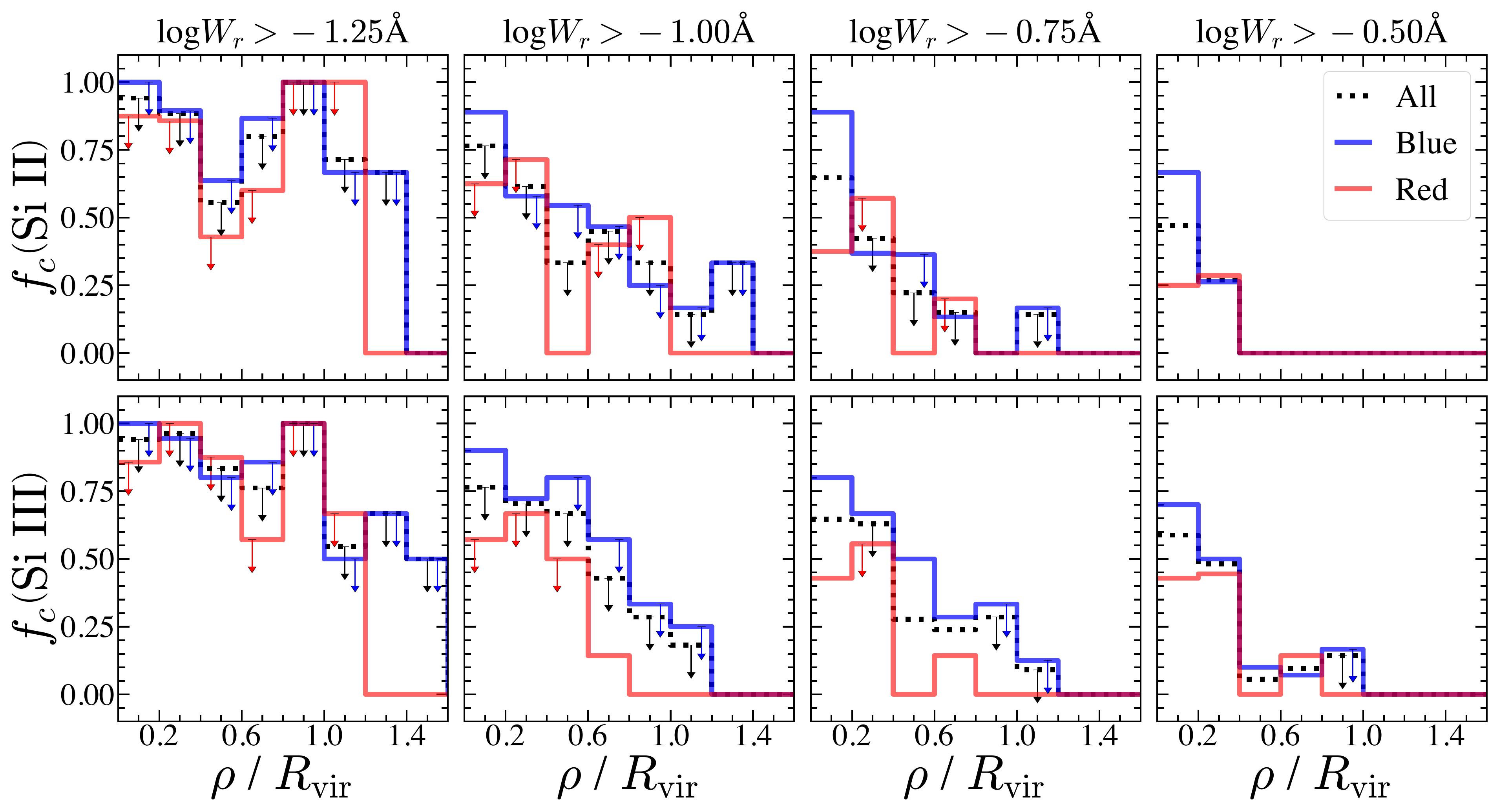}
    \caption{Covering fraction of \SiIIs (top) and \SiIIIs (bottom) as a function of impact parameter relative to the galaxy's virial radius for the combined COS-DIISC$+$COS-GASS$+$COS-Halos sample. Each column uses a different log rest-frame equivalent width limit, which is shown at the top of the column. The star-forming galaxies are shown in blue, passive galaxies in red, and all galaxies in black. Bins that include upper limit rest-frame equivalent widths are shown with a downward arrow.}
    \label{fig:cover_frac}
\end{figure*}

We show the covering fraction ($f_c$) of \SiIIs and \SiIIIs as a function of $\rho$ normalized by $R_\mathrm{vir}$ in Figure~\ref{fig:cover_frac}.
These were calculated for the combined COS-DIISC$+$COS-GASS$+$COS-Halos sample.
Various log$W_r$ cuts were used to look for trends with absorber strengths.
Bins that include upper limit measurements of log$W_r$ are shown with a downward arrow.
Radial declines are seen in most cases, with star-forming galaxies typically having higher $f_c$ values than passive galaxies.
This has been also been seen in simulations (e.g., \citealt{Appleby2021}).
As the log$W_r$ cut is increased, we see the decline in $f_c$ become steeper for both ions, which further indicates that the strongest absorbers tend to be found closer to galaxies.
The high values within $R_\mathrm{vir}$ are similar to what was found by other studies (e.g., \citealt{Werk2013,Liang2014,Richter2016,Lehner2020}).

\begin{figure*}
    \centering
    \includegraphics[width=\linewidth]{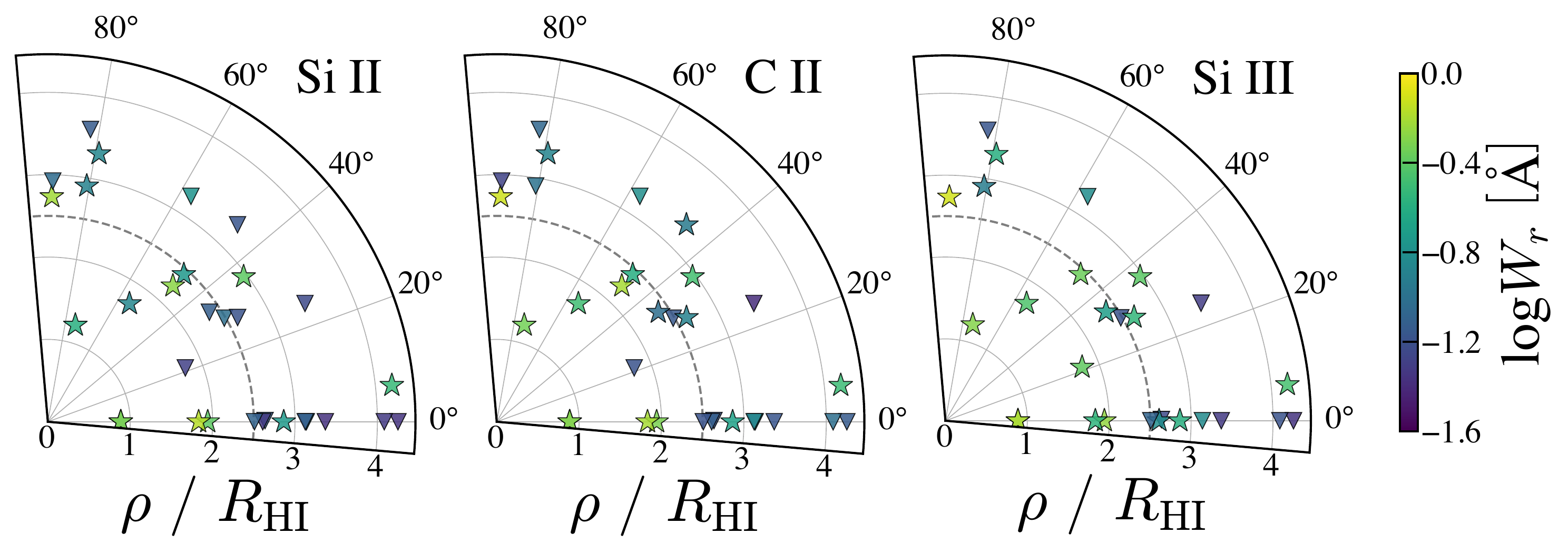}
    \caption{Position of QSO in terms of angle from the target galaxy's major axis and impact parameter relative to the galaxy's \HIs radius. Data points are colored based on the measured rest-frame equivalent width of \SiIIs (left), \CIIs (center), and \SiIIIs (right). Sight lines where the ion is detected are shown as stars, while those with nondetections are shown as downward triangles. \SiIIIs is detected in all sight lines within 2.5~$R_\mathrm{HI}$ as indicated by the dashed line and is seen in Figure~\ref{fig:Wr_rho_RHI_Rpetr}.}
    \label{fig:orient}
\end{figure*}

In Figure~\ref{fig:orient}, we show the measured log$W_r$ of \SiIIs (left), \CIIs (center), and \SiIIIs (right) as a function of $\rho$ normalized by $R_\mathrm{HI}$ and the angle from the galaxy's major axis.
We define galaxies with minor axis to major axis ratios greater than 0.8 to be face-on and these sight lines are considered to be passing through the extended galactic disk.
As a result, we set the orientation angle to $0\degree$ in the figure.
As previously noted, stronger components of each ion are found closer to the disk of galaxies. 
However, no trends are seen between the orientation of the QSO relative to the galaxy's major axis.

\subsection{Ionization State} \label{sect:ionstate}

Comparing the $N$ ratio of metals of different gas phase can reveal how ionized the gas is.
Here, we analyze the ratios of the silicon ions we have access to (i.e., $N(\mathrm{Si\ II})/N(\mathrm{Si\ III})$ and $N(\mathrm{Si\ IV})/N(\mathrm{Si\ III})$).
While other metal line ratios can be used, such as $N(\mathrm{Si\ III})/N(\mathrm{C\ II})$, we only analyze ions of silicon to limit the impact of dust depletion.

Most components had $N(\mathrm{Si\ II})/N(\mathrm{Si\ III})$ and $N(\mathrm{Si\ IV})/N(\mathrm{Si\ III})$ ratios near 1, though there is a rather large spread in each.
No trends were seen between either ratio and the properties of the host galaxy in Table~\ref{tab:sample}, suggesting that the galaxy itself has little impact on how ionized the CGM is.
We note, however, that the galaxies in the DIISC survey are typical $L_\star$ galaxies, and so trends may appear when looking at very small $\rho$ or around more extreme galaxies, such as starbursts.
Below, we explore processes that could be driving the observed ionization states.

\subsection{Ionization Equilibrium Models} \label{sect:ion_models}

The processes responsible for ionizing the CGM can be understood by comparing observed log$N$ values to those predicted from ionization models.
To determine if our absorbers are in photoionization equilibrium (PIE), we compare our measurements to \texttt{CLOUDY} models (v.23;~\citealt{Chatzikos2023}).
We assume the ions we have access to fall into three phases: a cold phase traced by \OIs and \SiII, a cool phase traced by \CIIs and \SiIII, and a warm phase traced by \SiIVs and \NV.
This gives us 19 cold phase components, 25 cool phase components, and 12 warm phase components to include in this analysis.
The results presented here are intended to be an initial constraint on the prevalence of PIE at the disk-CGM interface.
A more detailed analysis will be performed in a future study.

\begin{figure}
    \centering
    \includegraphics[width=\linewidth]{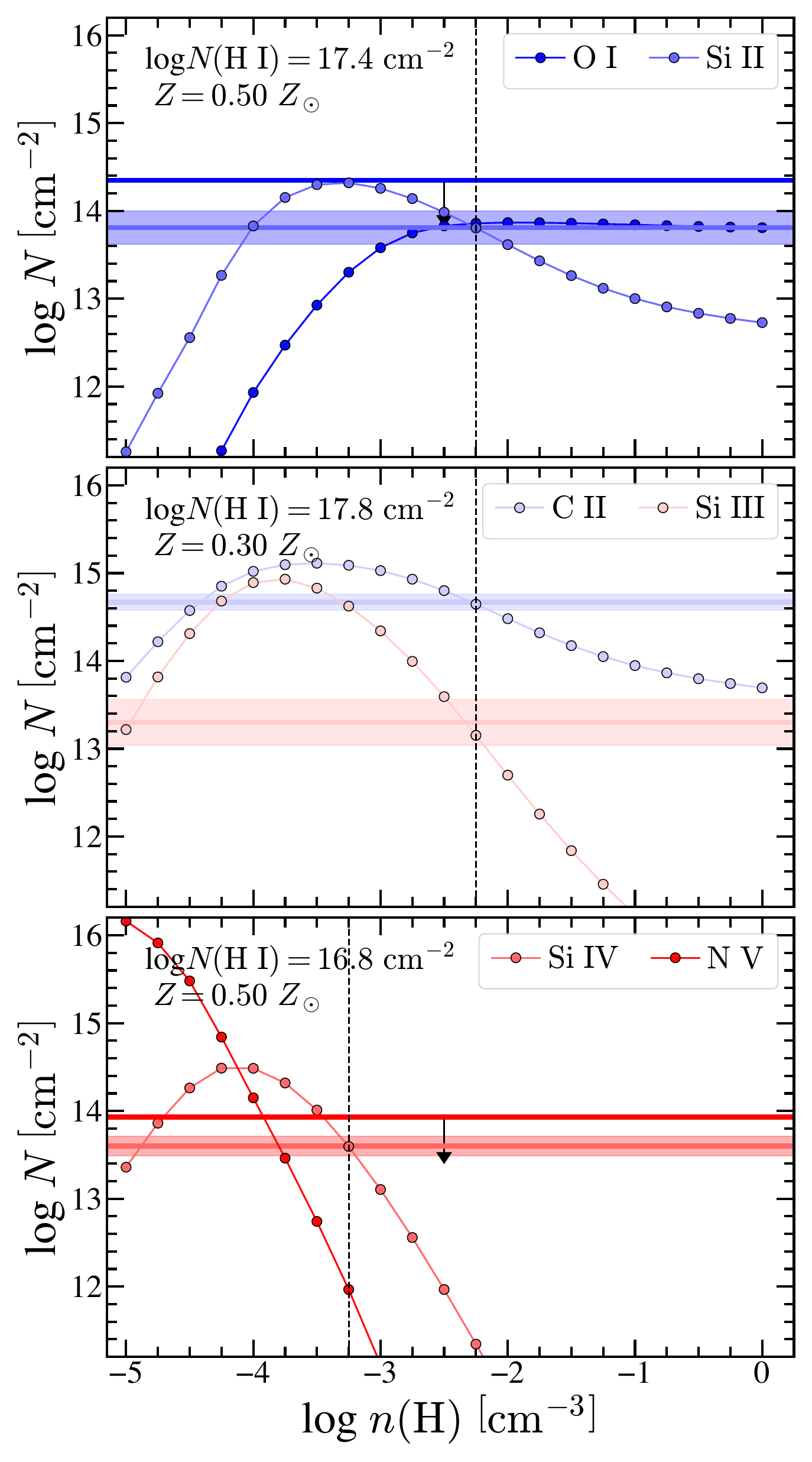}
    \caption{An example comparison of our measurements for the QSO-galaxy pair J1042+2501-NGC~3344 to \texttt{CLOUDY} PIE models. The top, middle, and bottom panels correspond to the cold, cool, and warm gas phases, respectively. Our measurements are shown as colored horizontal lines, with the shaded regions indicating the uncertainties. The color indicates the ion being plotted as shown in the upper right corner. The \HIs column density and metallicity of the model are shown in the upper left corner. The total H density that best reproduced the measurements is shown as vertical black dashed lines. Similar figures for each system included in this analysis are available in the online journal.}
    \label{fig:cloudy_example}
\end{figure}

Each \texttt{CLOUDY} model is exposed to a \citet{HM2012} extragalactic UV background as well as the Milky Way radiation field.
We use the Milky Way radiation fields presented in \citet{Fox2005} which are generated at $\rho =$ 10, 50, and 100~kpc.
For each sight line, we select the radiation field with the $\rho$ value that is closest to the one listed in Column (7) of Table~\ref{tab:sample}.
The total H density of the \texttt{CLOUDY} models was varied from log$\left[n(\mathrm{H})/\mathrm{cm}^{3}\right]$ = $-5.00$ to $0.00$ in steps of 0.25~dex.
A robust measure of log$N(\mathrm{H\ I})$ was difficult for most sight lines since the \Lyas features were saturated and we do not have access to other lines in the Lyman series.
As a result, we ran \texttt{CLOUDY} over a range of values for log$N(\mathrm{H\ I})$ in steps of 0.2 dex based on the measurements presented in \citet{Borthakur2024}.
Similarly, the metallicity along our sight lines is poorly constrained due to the lack of multiple Lyman series lines, leading us to assume metallicities of 0.04, 0.30, 0.50, and 1.00 times the solar value \citep{Asplund2009} motivated by other CGM surveys \citep{Lehner2013,Prochaska2017}.
Table~\ref{tab:cloudy_in} presents the input parameters used in our \texttt{CLOUDY} modeling.
This analysis is shown in Figure~\ref{fig:cloudy_example} for the QSO-galaxy pair J1042+2501-NGC~3344 as an example.
Similar figures are available for each sight line included in this analysis in the online journal.

\begin{deluxetable*}{cccccccC}
\centering
\tablecolumns{8}
\label{tab:cloudy_in}
\tablecaption{Input Parameters of \texttt{CLOUDY} PIE Models}
\tablehead{
\colhead{QSO} & \colhead{Galaxy} & \colhead{$z_\mathrm{sys}$} & \colhead{Cold} & \colhead{Cool} & \colhead{Warm} & \colhead{$\rho_\mathrm{MW}$} & \colhead{log$N(\mathrm{H\ I})$} \\
\colhead{(1)} & \colhead{(2)} & \colhead{(3)} & \colhead{(4)} & \colhead{(5)} & \colhead{(6)} & \colhead{(7)} & \colhead{(8)}}
\startdata
J0832+2431 & KUG 0829+246B & 0.0430 & 2 & 1 & 0 & 50 & 14.0 - 18.0  \\
J0835+2459 & NGC 2611 & 0.0175 & 2 & 3 & 3 & 50 & 15.0 - 19.8  \\
J0917+2719 & J0917+2720 & 0.0469 & 1 & 2 & 0 & 50 & 15.0 - 18.0 \\
J1042+2501 & NGC 3344 & 0.0020 & 1 & 1 & 1 & 50 & 14.0 - 18.0  \\
J1043+1151 & M95/NGC 3351 & 0.0026 & 1 & 1 & 1 & 10 & 14.0 - 18.0 \\
J1124+1131 & NGC 3666 & 0.0035 & 1 & 2 & 1 & 50 & 14.0 - 18.0 \\
J1133+2423 & NGC 3728 & 0.0232 & 2 & 2 & 0 & 100 & 14.0 - 18.0  \\
J1140+1136 & NGC 3810 & 0.0033 & 2 & 1 & 1 & 50 & 13.0 - 18.0  \\
J1157+0906 & AGC 213643 & 0.0367 & 0 & 1 & 1 & 100 & 14.0 - 18.0  \\
J1157+1140 & 2MASXJ1157+1139 & 0.0213 & 0 & 1 & 1 & 50 & 14.0 - 18.0  \\
J1221+0430 & M61/NGC 4303 & 0.0052 & 2 & 5 & 3 & 50 & 14.0 - 20.0  \\
J1314+2606 & J1314+2605 & 0.0431 & 1 & 1 & 0 & 100 & 14.0 - 18.0  \\
J1432+0955 & NGC 5669 & 0.0046 & 1 & 0 & 0 & 10 & 14.0 - 20.0  \\
J1524+0419 & 2MASXJ1524+0421 & 0.0374 & 1 & 2 & 0 & 100 & 14.0 - 18.0  \\
J1533+1501 & LEDA 140527 & 0.0429 & 1 & 1 & 0 & 10 & 15.0 - 21.0  \\
J1558+1205 & IC 1149 & 0.0156 & 1 & 1 & 0 & 50 & 14.0 - 18.0  \\
\enddata
\tablenotetext{}{Columns (1) and (2) are the QSO and galaxy, respectively, of the sight line. Column (3) is the redshift of the system. 
Columns (4), (5), and (6) are the number of cold, cool, and warm clouds, respectively, included in the analysis.
Columns (7) and (8) are the $\rho$ of the Milky Way radiation field and
the range of log$N(\mathrm{H\ I})$ used in the analysis.}
\end{deluxetable*}

It is important to highlight the limitations of this analysis.
Even though the CGM has been shown to contain multiple gas phases (e.g., \citealt{Werk2013,Cooper2021,Haislmaier2021,Sameer2024}), a clear delineation between the phases, as we are presenting, is unrealistic. 
A single ion, such as \SiIIIs or \SiIV, could be produced by multiple phases, which we do not account for here.
The Milky Way radiation fields included in the models is a constant source of photons, but this could vary with time.
While we are assuming solar relative abundances \citep{Asplund2009}, there is evidence that these can vary in galaxies and their CGM (e.g., \citealt{Fox2004,Izotov2023}).
It is known that some metals in the CGM of galaxies may be locked in dust and molecules; however, we are assuming there is no depletion in our measurements for this analysis.
As shown in Figure~\ref{fig:logN_b}, the largest absorbers, particularly of \CIIs and \SiIII, are saturated and could be the result of components overlapping in velocity space.
The impact all of these have on the results presented below will be explored in a future study.

We present the results of this analysis in the online journal. 
The majority of these absorbers (17 of 19 cold phase clouds, 22 of 25 cool phase clouds, and 10 of 12 warm phase clouds) are consistent with PIE models though we cannot rule out photoionization taking place in the remaining systems.
The cold components which are consistent with PIE were found to have log$\left[n(\mathrm{H})/\mathrm{cm}^{3}\right]$ between $-2.50$ and $0.00$ with temperatures between $T \approx 10^{3.5}$ and $10^{4.0}$~K.
The cool components were found to have log$\left[n(\mathrm{H})/\mathrm{cm}^{3}\right]$ between $-3.75$ and $-2.25$ with temperatures between $T \approx 10^{3.8}$ and $10^{4.1}$~K.
The warm components had log$\left[n(\mathrm{H})/\mathrm{cm}^{3}\right]$ between $-4.50$ and $-2.75$ with temperatures between $T \approx 10^{3.9}$ and $10^{4.1}$~K.

\begin{figure}
    \centering
    \includegraphics[width=\linewidth]{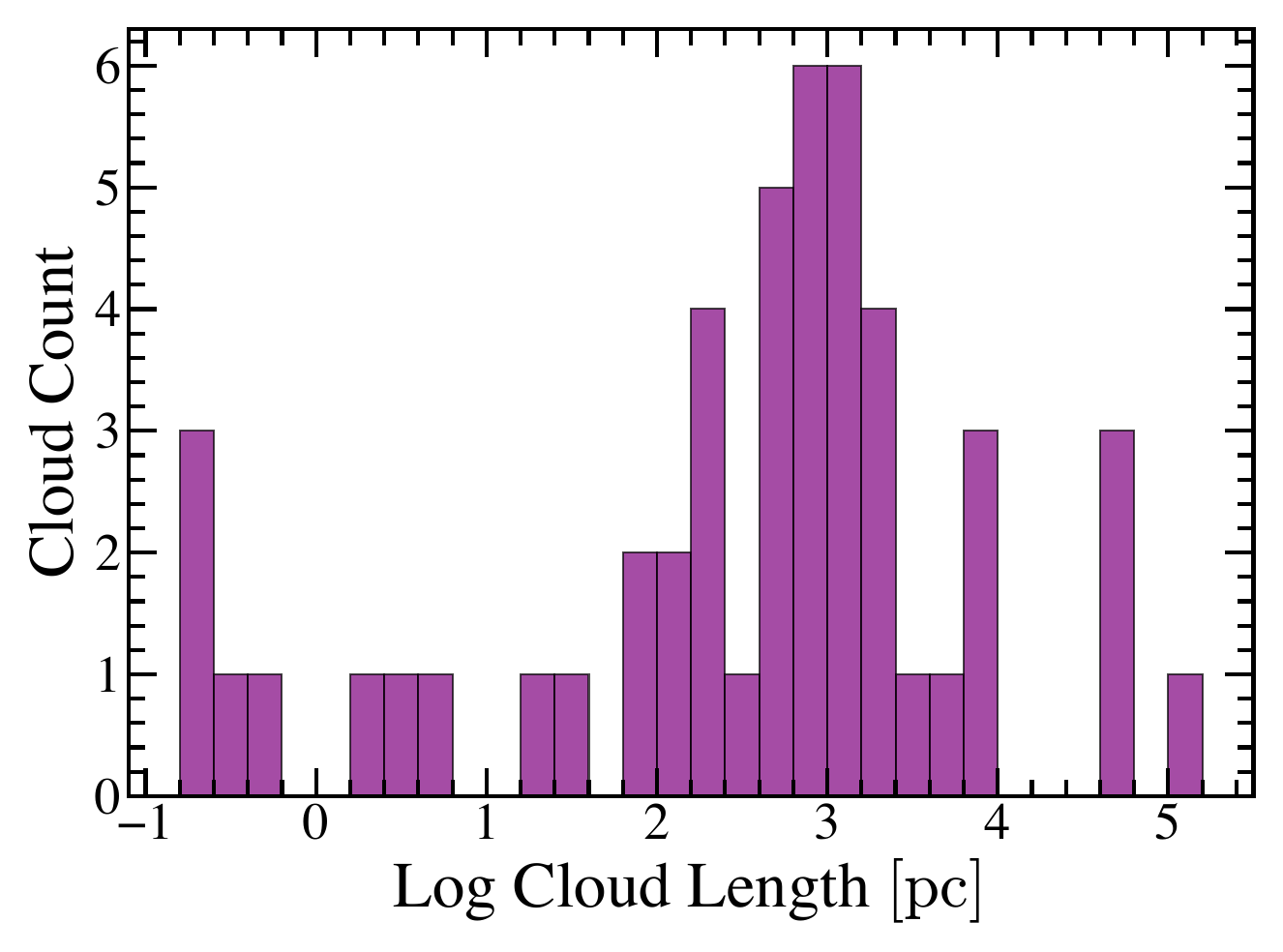}
    \caption{Derived cloud lengths based on \texttt{CLOUDY} models.}
    \label{fig:cloud_sizes}
\end{figure}

We show the distribution of the cloud lengths in Figure~\ref{fig:cloud_sizes}.
The cloud length can be determines as $N(\mathrm{H})~/~n(\mathrm{H})$.
We use the $N(\mathrm{H})$ from the model, such that $N(\mathrm{H}) = N(\mathrm{H\ I}) + N(\mathrm{H\ II})$.
All of the cold phase clouds were found to be $\lesssim$1~kpc in length.
The cool phase clouds were all smaller than 7~kpc in length though most (19 / 22) were smaller than 2~kpc.
Meanwhile, all of the warm phase clouds were larger than 1~kpc though the majority (6 / 10) are smaller than 10~kpc.
These sizes should be thought of as upper limits given that at least some of the absorbers with the largest $N$ are likely comprised of many smaller clouds overlapping in velocity space and we do not have the spectral resolution to resolve each cloud individually.
These sizes are similar to the clouds derived by the CUBS collaboration \citep{Zahedy2021} as well as those inferred by \citet{Keeney2017} but are smaller than those from the COS-Halos survey \citep{Werk2014,Werk2016}.
This consistency between surveys suggest these lengths are characteristic cloud lengths in the CGM regardless of the phase being probed.
As discussed above, the analysis presented is intended to be a first pass at modeling our measurement. 
A more detailed study will be implemented in a future study.

\begin{figure}
    \centering
    \includegraphics[width=\linewidth]{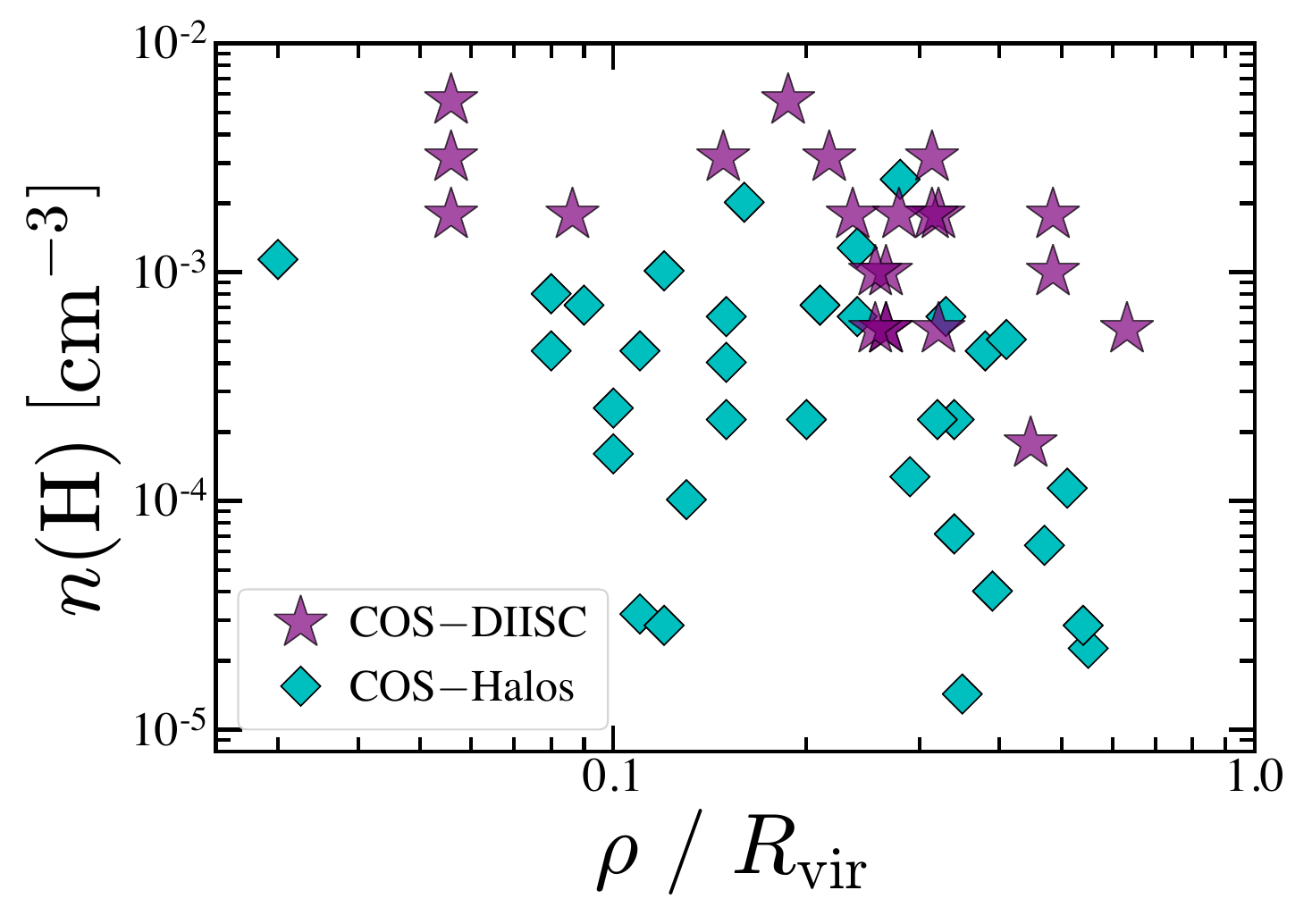}
    \caption{Inferred total hydrogen density as a function of impact parameter normalized by the galaxies virial radius. Values from the cool clouds in our sample (COS-DIISC) are shown as purple stars, while those from the COS-Halos sample are shown as cyan diamonds.}
    \label{fig:nH_rRvir}
\end{figure}

In Figure~\ref{fig:nH_rRvir} we plot the $n(\mathrm{H})$ values inferred from the \texttt{CLOUDY} models for our cool clouds as a function of $\rho$ normalized by $R_\mathrm{vir}$.
We also plot the values from COS-Halos \citep{Werk2014} for comparison.
We only show the results for the cool clouds to better compare to the COS-Halos sample.
Both of these samples show a similar distribution.

\begin{figure}
    \centering
    \includegraphics[width=\linewidth]{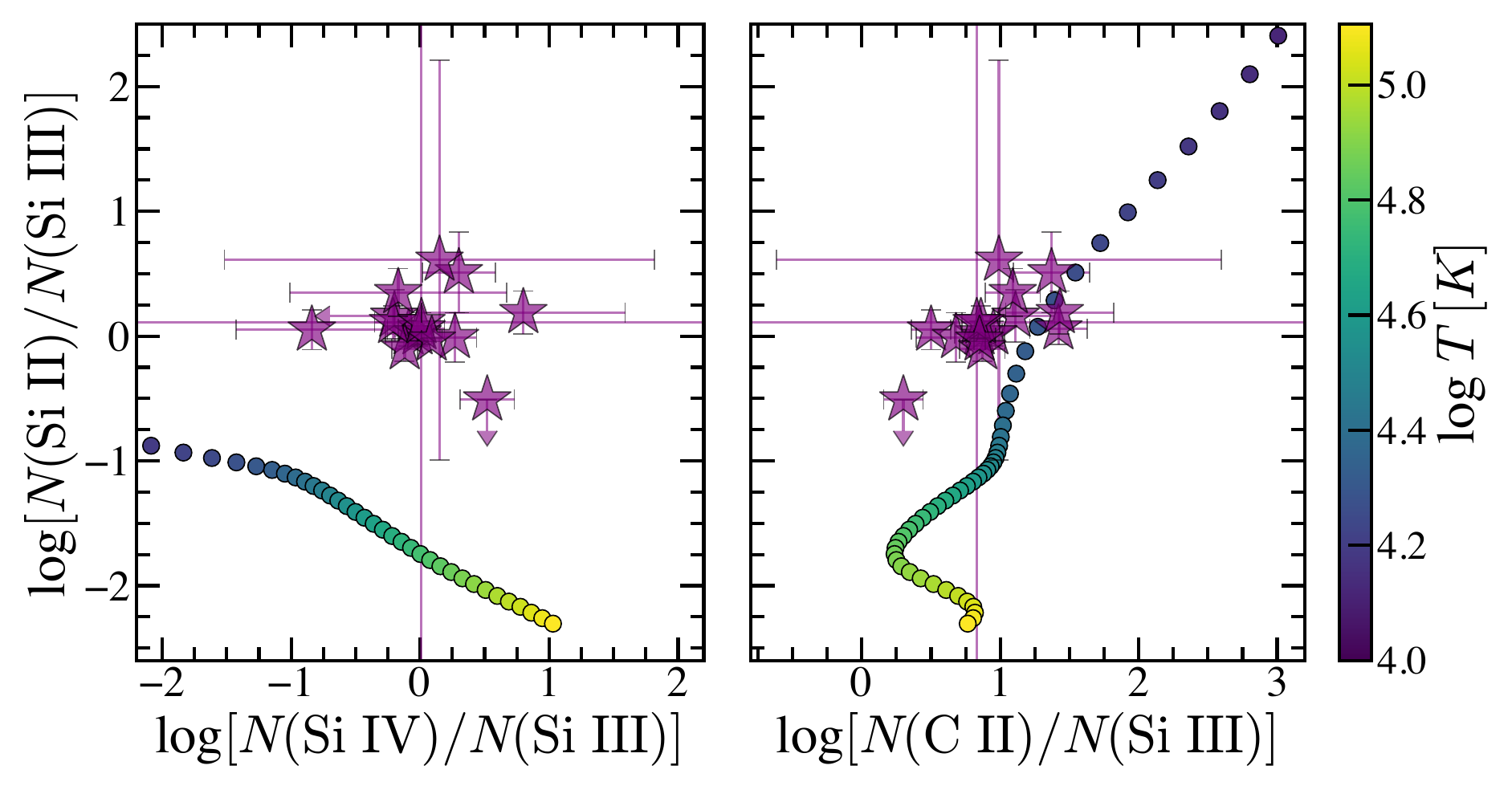}
    \caption{Comparing CIE models from \citet{Gnat2007}, colored circles, to observed column density ratios, stars. The color of the circles corresponds to the temperature of the CIE model.}
    \label{fig:cie}
\end{figure}

To determine whether the components along our sight lines are in collisional ionization equilibrium (CIE), we compare our metal line ratios to the solar metallicity models from \citet{Gnat2007}, which are shown in Figure~\ref{fig:cie}.
The left panel compares the different ionization states of silicon we have access to, which covers a range of IPs (16.4 $-$ 45.1~eV), while the right panel probes more coherent phases covering a narrower range of IPs (16.4 $-$ 33.5~eV).
In the right panel, many of our points are consistent with CIE at $T \approx 10^{4.4}-10^{4.5}$~K; however, they do not match our measurements shown in the left panel of the figure.
These show that low-ion components are likely not in CIE though we are not able to conclusively rule this out.
Though the high-ion components of \SiIVs and \CIVs could be in CIE since there can be different ionization mechanisms for different phases of gas.
Similar to the PIE analysis above, this is intended to be an initial constraint on whether our measurements could be explained using CIE models.
We will explore these models in more detail in a future study to robustly determine the role collisional ionization plays in our measurements.

\subsection{Stacking Spectra}

\begin{figure*}
    \centering
    \includegraphics[width=\linewidth]{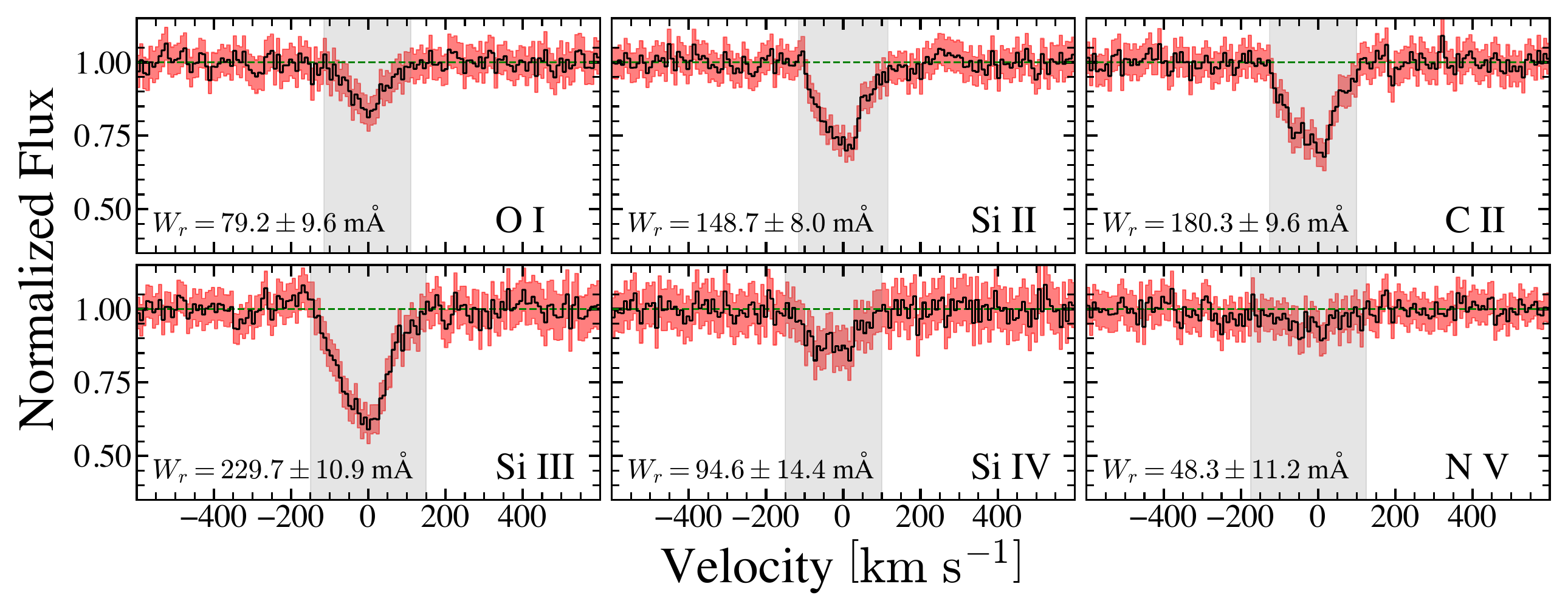}
    \caption{Stacked spectra of all sight lines at the velocity of each ion. The metal being shown is indicated in the bottom right of each panel. The 1$\sigma$ uncertainties are shown as red shaded regions. The gray regions highlight the velocity range used to calculate the rest-frame equivalent width of the spectra, which we show in the bottom left corner in units of m\AA.}
    \label{fig:stack}
\end{figure*}

To study the average properties of the CGMs in our sample rather than individual systems, we stack our spectra at the $v_\mathrm{sys}$ of the host galaxy.
For ions with multiple transitions, we only stack the strongest transitions.
The results of this stacking is shown in Figure~\ref{fig:stack}.
All of the ions show absorption features above our 3$\sigma$ detection limit.

\SiIIIs has the largest absorption, as measured by $W_r$, of any ion, followed by \CII.
The depth of the stacked absorbers are independent constraints on the $f_c$ of ions.
The \OI, \SiIV, and \NVs spectra are relatively shallow, leading to $f_c < 20$\%.
\SiII, \CII, and \SiIII, meanwhile, had $f_c$ between 30\% and 41\%, similar to what was found at $\sim$$R_\mathrm{vir}$ in Figure~\ref{fig:cover_frac}.

We constrained the $v_\mathrm{obs}$ of these stacked spectra using the apparent optical depth method \citep{Savage1991}.
The $v_\mathrm{obs}$ of each stack is within $\pm$25~\kms, indicating that most of the absorption is occurring at or near the galaxy's $v_\mathrm{sys}$.

These trends suggest that the most prominent gas phase, of those we have access to, in the CGM at low $z$ is the cool-warm phase, as traced by \CIIs and \SiIII, and that most of the absorption is taking place at a similar velocity as the host galaxy.
However, it is important to note that the $z$ of our sample galaxies prevent us from studying metals with higher IPs such as \ion{O}{6} and \ion{Ne}{8} with COS.

\section{Summary \& Conclusions} \label{sect:summary}

Here, we have investigate the content, kinematics, and ionization of UV metal absorption lines in the CGM surrounding 31 galaxies ($z_\mathrm{sys} = 0.0020 - 0.0507$) from the DIISC survey using HST/COS spectra.
The selection criteria of the DIISC survey allow us to determine the impact that a galaxy's \HIs disk has on its CGM as never before.
The main conclusions of our analysis are as follows:

\begin{enumerate}
    \item Of the ions under study, \SiIIIs was detected most frequently along the sight lines, with 18 of 31 galaxies ($\sim$55\%) showing at least one component. The vast majority of these (13;~$\sim$77\%) showed multicomponent structures. All ions showed fewer components per sight line on average when probing the CGM at larger $\rho$. 
    
    \item We find stronger absorbers (i.e., larger log$W_r$) of each ion closer to the disk of galaxies. The majority of sight lines within $\sim$2.5~$R_\mathrm{HI}$ contained metals, while those beyond this were more likely to be nondetections. These results indicate that the radial profile of the metal ions with IPs between 13.6~eV and 33.5~eV are more correlated with $\rho$~/~$R_\mathrm{HI}$, rather than the $\rho$~/~$R_\mathrm{vir}$ measure typically used by CGM surveys.

    \item In most ions, a strong correlation is seen between the Doppler $b$ parameters and log$N$ of individual components. When taken with the above points, this trend can be explained by many unresolved, overlapping components appearing as one large component near the galaxy's $v_\mathrm{sys}$ suggesting clustering of clouds close to the galaxy's $v_\mathrm{sys}$. 

    \item Many of our components were consistent with PIE models from \texttt{CLOUDY} regardless of the gas phase being probed. This suggests that photoionization is the dominate ionization process near the disk-CGM interface. These models reveal that our components probing the cold gas phase are produced by clouds with lengths $\lesssim$1~kpc. The cool gas phase was largely produced by clouds smaller than 2~kpc in length. Meanwhile, the majority of the components probing the warm gas phase were smaller than 10~kpc in length. No low-ion components were found to be consistent with CIE models though we cannot rule out the high-ion absorbers being in CIE. It is important to note that this analysis is intended to be a first pass at modeling our measurements using PIE and CIE models. A more detailed analysis of these will be explored in a future study.
\end{enumerate}

These results allow us to further study the distribution of CGM gas around low-$z$ galaxies and constrain models that can reproduce the trends.
We will explore this fully in a future publication.

\begin{acknowledgments}
We thank the referee for constructive comments, which helped improve the robustness and clarity of the manuscript.
B.K., S.B., M.P., T.M., and T.H. are supported by HST grant HST-GO-14071 administrated by STScI, which is operated by AURA under contract NAS 5- 26555 from NASA and the NSF grants 2108159 and 2009409. S.B., and M.P. are supported by NASA ADAP grant 80NSSC21K0643.

B.K. would like to thank the Oases in the Cosmic Desert 2023 CGM conference for inspiring some of the analysis presented in this work.
B.K., S.B., M.P., and T.M. acknowledge the native people and the land that Arizona State University’s campuses are located in the Salt River Valley.
The ancestral territories of Indigenous peoples, including the Akimel O’odham (Pima) and Pee Posh (Maricopa) Indian Communities, whose care and keeping of these lands allows us to be here today.
\end{acknowledgments}

\facilities{HST (COS); SDSS.}

\software{\texttt{matplotlib} (v3.2.2;~\citealt{Hunter2007}), \texttt{astropy} (v4.2.1;~\citealt{Astropy2013,Astropy2018}), \texttt{CLOUDY} (v.23;~\citealt{Chatzikos2023}), \texttt{numpy} (v1.22.0;~\citealt{Harris2020}), \texttt{scipy} (v1.6.2;~\citealt{Virtanen2020}), \texttt{pandas} (v1.3.5;~\citealt{Reback2021}).}

\bibliographystyle{aasjournal}
\bibliography{References}

\end{document}